\documentclass[aps,floats,twocolumn,epsf,prb,showpacs]{revtex4-1}
\usepackage{graphicx}
\usepackage{epstopdf}

\usepackage{bm}

\usepackage{color}

\newcommand{\nop}[1]{}

\begin{document}


\title{Divergences of the irreducible vertex functions in correlated
  metallic systems: \\ Insights from the Anderson impurity model}

\author{P. Chalupa$^a$, P. Gunacker$^a$, T. Sch\"afer$^{a,b,c}$, K. Held$^{a}$, and A. Toschi$^a$}

\affiliation{$^a$ Institute of Solid State Physics, Technische Universit\"at Wien, 1040 Vienna, Austria}
\affiliation{$^b$  Coll\'ege de France, 11 place Marcelin Berthelot, 75005 Paris, France}
\affiliation{$^c$  Centre de Physique Th\'eorique, CNRS, \'Ecole Polytechnique, 91128 Palaiseau, France}

\date{ \today }

\begin{abstract}
In this work, we analyze in detail the occurrence of divergences in the irreducible vertex functions for one 
of the fundamental models of many-body physics: the Anderson impurity model (AIM). These 
divergences, a 
surprising hallmark of the breakdown of many-electron perturbation 
theory -- have been recently observed in 
several contexts, including the dynamical mean-field solution of the Hubbard model. The numerical calculations for the AIM presented in this work, as well as their 
comparison with the corresponding results for the Hubbard model, allow us to clarify several open questions 
about the properties of vertex divergences in a 
particularly interesting context, the 
correlated metallic regime at low-temperatures. Specifically, our analysis (i) rules out explicitly 
the transition to a Mott insulating phase, but not the more general 
suppression of charge fluctuations (proposed in [O. Gunnarsson {\it et al.}, Phys.\,Rev.\,B {\bf 93},\,245102\,(2016)]), 
as a necessary condition 
for the occurrence of 
vertex divergences, (ii)  
clarifies their relation with the underlying Kondo physics, and, eventually, (iii) individuates which 
divergences might also appear on the real frequency axis in the limit of zero temperature, 
through the discovered scaling properties of the singular eigenvectors.

\end{abstract}

\maketitle

\let\n=\nu \let\o =\omega \let\s=\sigma


\section{Introduction}
\label{sec:intro}

The foundation of the Feynman diagrammatic technique relies on the 
many-body perturbation expansion. Nonetheless, its high 
flexibility and the transparency of the physical interpretation have
often motivated the application of diagrammatic schemes also well \emph{beyond} the perturbative regime. 
This is particularly true for the many-electron problem in condensed
matter theory.  In fact, for the latter, the
identification of a small parameter controlling the perturbation
expansion can become a very hard task, especially if 
the Coulomb interaction is not sufficiently screened, a common situation
in transition metal oxides and heavy fermion compounds.

In general, exploiting  diagrammatic techniques
beyond the regime of validity of the underlying
perturbation expansion is a viable option, and -- in some cases
-- also a rewarding one, as witnessed, e.g., by the success of the Dynamical
Mean-Field Theory (DMFT)\cite{DMFTREV} and its 
extensions\cite{DCAREV,Rohringer_REV}. 
However, in doing so, one must expect to face particular
problems, which might limit the applicability of  well known
diagrammatic relations and challenge the corresponding algorithmic
implementations in the strong-coupling regime. 
Not surprisingly, considering the fast developments of the diagrammatic
extensions\cite{DGA, DF, DB, 1PI, DMF2RG, TRILEX, DMFT_FLEX, QUADRILEX, Rohringer_REV} of 
DMFT, some of
these issues have been recently put in the focus of the forefront
literature on quantum many-body theory. 

In particular, two main kind of problems have been 
brought to light\cite{Schaefer2013, Kozik2015} and analyzed. 
The first one  is the occurrence of \emph{divergences} of the two-particle
irreducible (2PI) vertex functions
in several many-electron models. In fact, their occurrence has been
reported, even for moderate values of the electronic interaction, in
DMFT studies of the disordered binary mixture (BM)\cite{Schaefer2016}, the
Falicov Kimball (FK)\cite{Schaefer2013,Schaefer2016,Janis2014,Ribic2016} and the 
Hubbard model\cite{Schaefer2013,Schaefer2016,Rohringer_Diss, Vucevia2018}. Analytical
calculations\cite{Schaefer2013,Schaefer2016,Rohringer_Diss,Stan2015}
for the atomic limit (AL) of the
Hubbard model\cite{Hubbard} or the one point model\cite{Lani2012, Berger2014} have provided further evidence of
the robustness and the generality of the occurrence of these 
divergences of the 2PI vertex.
Finally, calculations with the dynamical cluster
approximation (DCA) have also demonstrated\cite{Gunnarsson2016, Vucevia2018}  that the
observed divergences are not an artifact of the purely local treatment
of DMFT. 
The irreducible vertex divergences appear as a 
consequence of the non-invertibility of the Bethe-Salpeter equation in the fermionic  Matsubara 
frequency space (see Sec.\,II), associated with a
simultaneous\cite{Schaefer2016} non-invertibility of the parquet equations.
In specific cases (FK\cite{Janis2014,Ribic2016,Schaefer2016}, BM\cite{Schaefer2016}), 
their presence has been also reported on the real frequency axis.

The second problem 
reported \cite{Kozik2015,Schaefer2016,Stan2015,Rossi2015,Keiter, Gunnarsson2017, Vucevia2018} 
is an intrinsic \emph{multivaluedness} of the
Luttinger Ward 
functional (LWF). This
unexpected characteristic of the LWF has been
demonstrated\cite{Kozik2015} by considering the self-consistent (bold) perturbation expansion for the 
self-energy
$\Sigma[G]$ in the AL of the Hubbard model, for which the exact Green's function is known
analytically. The corresponding resummation is found always to
converge but, for interaction values $U$ larger than a specific
$\widetilde{U}$, it converges to unphysical results, indicating the
existence of at least two branches of the LWF for the self-energy.  

Eventually, recent studies\cite{Gunnarsson2017} have rigorously demonstrated 
that these two aspects are exactly related, providing an analytic
proof that any crossing of different branches in the LWF
functional is associated to a divergence of the 2PI vertex, occurring 
for the same parameters (see Sec.\,A in the Supplemental Material of Ref.\,[\onlinecite{Gunnarsson2017}]). 

From a purely theoretical viewpoint, these problems can be regarded
as complementary manifestations of the breakdown of the perturbation
expansion. At the same time, 
from a more practical perspective, their potential impact on several cutting edge algorithmic
developments can also be significant. In particular, we recall that
the 2PI vertex functions constitute the fundamental building block of all diagrammatic
theories based on the Bethe-Salpeter or parquet equations\cite{Bickers,Yang2009,Tam2013,Li2017}, 
such as, e.g., the dynamical vertex approximation
(D$\Gamma$A)\cite{DGA,Toschi2011,Galler2017}, the multiscale\cite{Slezak2009}, and
the quadruply-irreducible local expansion (QUADRILEX)\cite{QUADRILEX} approach, or the parquet decomposition of
the self-energy\cite{Gunnarsson2016}.
Similarly, the presence of multiple branches in the LWF might
pose difficulties to bold diagrammatic Monte Carlo schemes\cite{Kozik2015,Tarantino2017,Vucevia2018}.  
The discussion on how (and to what extent) it is possible to circumvent
these difficulties within the different algorithms is a subject of 
current scientific debate\cite{note_wayouts}. 

One should mention, moreover, that the interpretation of the physics
underlying this twofold manifestation of the
breakdown of the perturbation expansion is still debated. Certainly, 
these crossings and divergences are \emph{not} associated with any thermodynamic
phase transition, due to the mutual compensation of  
(divergent) irreducible and fully irreducible diagrams in the parquet equations, which
ensures, that the full vertex stays finite 
(see Sec.\,\ref{sec:formalism}). It has been
proposed\cite{Schaefer2013,Janis2014,Schaefer2016}, however, that they might be
interpreted, for example, as precursors of the Mott metal-insulator
transition\cite{MH} (MIT), as features of the separation of spectral weight (such as the Hubbard subbands), 
as an implication of the
emergence of kinks in the spectral function\cite{Byzuk2007, Held2013} and
specific heat\cite{Toschi2009, Toschi2010}, and even in terms of qualitative changes in the
non-equilibrium asymptotic behaviors\cite{Schaefer2013}. 
Subsequently, it has been argued\cite{Gunnarsson2016, Gunnarsson2017}, that their occurrence in the 
Hubbard model is associated with the progressive suppression of the charge 
susceptibility for increasing values of the 
electrostatic repulsion $U$.

The aim of this paper is to improve our current understanding of the properties of the
2PI-vertex divergences, especially in the arguably most
interesting parameter regime of low-temperatures and moderate interaction
values, where they appear to coexist with a metallic, Fermi-Liquid
ground state.   

This will provide, in turn, hitherto missing pieces of
information about the vertex divergences. In fact, 
recent studies\cite{Schaefer2016,Janis2014,Ribic2016} have reported
progress in understanding  the (relatively) simpler region of high temperatures and
large interactions: Here, the properties of the DMFT vertex functions
of the Hubbard model are efficiently approximated by easier calculations performed on the 
one hand, in the BM and 
FK\cite{Janis2014,Ribic2016,Schaefer2016} case, whose DMFT solutions 
correspond\cite{SchwartzSiggia1972,Vlaming1992} essentially to the 
coherent potential approximation (CPA)\cite{CPA_Rev}, and on the other hand 
in the AL\cite{Schaefer2016} case. In these models, it has been
shown\cite{Schaefer2016} that the proliferation of the divergence
lines in the corresponding phase-diagrams is merely a consequence of the Matsubara
representation of a unique underlying
energy scale $\nu^*$, which, in the case of the BM, completely controls all the vertex divergences. As a result,
all the divergence lines in the phase diagrams of the BM collapse
onto a single one, if multiplied with the appropriate Matsubara index
($2n-1$). 
For the FK and the AL case this precise characterization applies, however, only to 
half of the divergence lines\cite{Schaefer2016,Rohringer_AL}.
Further, at $T \rightarrow 0$ all the lines accumulate at the
same value of $U$, where the vertex is found to be diverging even on
the real frequency axis. Eventually, the scale $\nu^*$ could be directly related
to specific properties which characterize  the single particle Green's
function of the model evolving towards the opening of a Mott spectral
gap. More precisely in the BM and FK model $\nu^*$ corresponds to the frequency where the 
minimum of $\mathrm{Im}\, G(i\nu_n)$ is found\cite{Schaefer2016}, 
in the AL, interestingly, this is the case for the 
inflection point\cite{Rohringer_AL}.

None of these semi-analytical results, however,  turned out to be applicable
for the interpretation of the low-$T$ vertex divergences in the Hubbard model.
In fact, in the low-$T$ region of the corresponding DMFT phase diagram, the 
divergence lines display a clear
re-entrance, somehow similarly shaped as the Mott-Hubbard MIT,
with a significant spread for $T \rightarrow 0$. Consequently, no
unique energy scale $\nu^*$ could be identified, no collapse of the
lines is observed, as well as any accumulation at a specific $U$
value for $T=0$.  Also, classifying the different
types of divergences according to their locality in frequency space\cite{Schaefer2016} (see also
Sec.\,\ref{sec:formalism}), 
appears to work no longer.
The  plausible origin of these complications w.r.t. FK or AL must lie
in the differences of the underlying physics. The major one is, arguably, the presence
of low-energy coherent quasiparticle excitations in the correlated
metallic region of the Hubbard model: These are missing -- per
construction -- in the simpler cases of FK and AL. 

The path towards a better understanding of the nature of the vertex
divergences in the correlated metallic regime is hampered
by the intrinsic feedback effects of the self-consistency procedure in DMFT: The embedding bath
of the auxiliary Anderson impurity model\cite{Anderson}, 
for which the vertex functions are computed, is
continuously readjusted, including in itself an important part of the correlation
features of the DMFT solution.  For example, it has been pointed
out\cite{Held2013}, that these self-consistent effects are responsible for
the appearance of two different low-energy scales ($\omega_{FL}$ and
$\omega_{CP}$ following the notation\cite{scales_note} of Ref.\,[\onlinecite{Byzuk2007}]) 
and, thus, for the related low-energy kinks in the
self-energy\cite{Byzuk2007} and the specific heat\cite{Toschi2009} . 

To avoid this additional complication, in this we will 
disentangle the different effects by considering
a more basic system than the Hubbard model, still capable, however, of
capturing the physics of low-energy quasiparticle excitations: the
Anderson impurity model (AIM). In fact, the AIM, defined by a
fixed electronic bath embedding one correlated impurity site, 
describes highly non-perturbative processes (such as the many-body effect 
related to the Kondo screening), but, at 
the same time, yields important simplifications of the underlying physics
w.r.t. the self-consistent solution of DMFT: For example, no
Mott-Hubbard MIT is present at T=0, so that the ground state properties 
remain Fermi liquid-like for all values of the local electrostatic repulsion $U$.
In particular, the comparison of our results for the vertex divergences of the AIM
to the ones found in the Hubbard model will allow us to rigorously address 
a set of important questions, left unanswered in the most recent literature:

\begin{description}

\item[(i)] Is the Mott-Hubbard MIT a necessary condition for the occurrence of 
the vertex divergences and their related manifestations? 

\item[(ii)] Given that for the Hubbard model at low-$T$ it was not possible to identify a
  unique scale $\nu^*$, can there be a scenario, comprising two-energy scales on the
  real-axis ($\omega_{FL}$ and $\omega_{CP}$) compatible with the
  vertex divergences? In the case of a positive answer, can
  one find a relation with the low-energy kink(s)\cite{Byzuk2007,Held2013} in the self-energy 
  and $C_V(T)$\cite{Toschi2009}, found in previous DMFT studies.
  However, for the AIM with large conduction electron bandwidth studied in this paper, 
  only one energy scale, the Kondo scale\cite{Kondo} $T_K$, exists. 
 
\item[(iii)] Can vertex divergences on the real frequency axis be expected,
 similarly as in the BM/FK case?

\item[(iv)] What is the role played by the Kondo scale\cite{Kondo}, which has
  -- for the case of the AIM -- a direct physical meaning?
  
\item[(v)] Can one exploit  the simpler AIM results presented in this work, to predict some still
  unknown aspects of the divergences in the Hubbard model?
  
\end{description}

The paper is structured as follows: In Sec.\,II we define the
specific AIM used in our calculations as well as the 
quantum field theory formalism necessary to analyze the irreducible vertex 
divergences, and describe concisely the numerical method applied as impurity solver. 
Thereafter, in Sec.\,III, our numerical results, together with a
comparison to previous DMFT findings for the Hubbard model, are
presented. In Sec.\,IV, a detailed analysis of the data shown in
Sec.\,III is made, providing clear-cut answers to the specific 
questions $\textbf{(i)}$-$\textbf{(v)}$ posed above.
Finally, in Sec.\,V, a conclusion and an outlook of our work are presented.  

\section{Formalism and Methods}

\subsection{Anderson impurity model}

In this work we consider an AIM with a fixed hybridization
to a bath of non-interacting electrons with a 
constant (box-shaped) density of states (DOS). The corresponding 
Hamiltonian reads as

\begin{eqnarray}
\label{equ:ham_AIM}
\mathcal{H} &=& \sum_{\sigma} \epsilon_d^{\phantom \dagger} 
d^{\dagger}_{\sigma} d^{\phantom \dagger}_{\sigma} + U n_{d,\uparrow} 
n_{d,\downarrow} + 
\\ 
\sum_{\mathbf{k},\sigma} &\epsilon^{\phantom \dagger}_{\mathbf{k}}& 
c^{\dagger}_{\mathbf{k},\sigma} 
c^{\phantom \dagger}_{\mathbf{k},\sigma}
+ \sum_{\mathbf{k},\sigma} \Big( V^{\phantom \dagger}_{\mathbf{k}} d^{\dagger}_{\sigma} 
c^{\phantom \dagger}_{\mathbf{k},\sigma} + 
 V^{\ast}_{\mathbf{k}} c^{\dagger}_{\mathbf{k},\sigma} d^{\phantom \dagger}_{\sigma} \Big) \, , \nonumber
\end{eqnarray}
where $\epsilon_d$ represents the energy of the impurity level, $U$ is the value of the 
local interaction 
and $d_{\sigma}^{\dagger}$/$d^{\phantom \dagger}_{\sigma}$ creates/annihilates 
an electron on the impurity site, $n_{d,\sigma} = d_{\sigma}^{\dagger}d^{\phantom \dagger}_{\sigma}$.
The first term in the second line of Eq.\,(\ref{equ:ham_AIM}) is the kinetic energy of the 
non-interacting bath of electrons with $\epsilon_{\mathbf{k}}$ as the dispersion relation and 
$c_{\mathbf{k}, \sigma}^{\dagger}$/$c^{\phantom \dagger}_{\mathbf{k}, \sigma}$, the 
creation/annihilation operators 
of the bath electrons. Finally, the last terms represent the hopping onto/off the impurity site.
In the specific AIM chosen for this work the DOS of the bath electrons is 
$\rho(\epsilon) = (1/2D) \Theta(D-|\epsilon|)$, with
the half-bandwidth $D=10$ 
being the largest energy scale of the system. 
The hybridization is assumed to be $\mathbf{k}$-independent and set to $2$ 
($V_\mathbf{k} = V = 2$) and the chemical potential is set to $\mu = U/2$ 
(half-filled/particle-hole symmetric case). The choice of a box-shaped DOS and a $\mathbf{k}$-independent 
hybridization ensures that no particular features of $\rho(\epsilon)$ or $V$ will 
affect the study of irreducible vertex divergences, and the selected parameter set should
guarantee, that the Kondo temperature of our AIM remains sizable with respect to the other energy 
scales, for the half-filled case considered. 

\subsection{Two-particle formalism}
\label{sec:formalism}

The two-particle irreducible vertex function $\Gamma$ , whose divergences will be studied in this work, 
is -- per definition -- the fundamental building block of the Bethe-Salpeter equation for the 
generalized susceptibility. While for more detailed information and definitions
we refer the reader to Ref.\,[\onlinecite{Rohringer2012}] as well as 
Refs.\,[\onlinecite{Rohringer_Diss,Schaefer2016,Rohringer_REV,Wentzell_frequencies_2016,Agnese_2018}], 
we want to summarize 
here solely the crucial objects necessary for our analysis. We 
start, then, from the generalized 
susceptibility $\chi_{ph,\sigma \sigma'}^{\nu_n \nu_{n'} \Omega_n}$ at 
the impurity site, which is defined (in the 
particle-hole 
channel) as 

\begin{eqnarray}
\label{equ:form_gen_chi}
\chi_{ph,\sigma \sigma'}^{\nu_n \nu_{n'} \Omega_n} &=& \int \limits_0^\beta d \tau_1 d\tau_2 d\tau_3 \,
e^{-i\nu_n \tau_1} 
e^{i(\nu_n + \Omega_n)\tau_2} e^{-i(\nu_{n'} + \Omega_n)\tau_3} \nonumber  \\
&\times & [ \langle T_{\tau} d_{\sigma}^{\dagger} (\tau_1)
d_{\sigma}^{\phantom \dagger}(\tau_2) d_{\sigma'}^{\dagger}(\tau_3) d_{\sigma'}^{\phantom \dagger}(0) 
\rangle \\
& -& \langle T_{\tau} d_{\sigma}^{\dagger} (\tau_1)
d_{\sigma}^{\phantom \dagger}(\tau_2) \rangle \langle T_{\tau} d_{\sigma'}^{\dagger}(\tau_3) 
d_{\sigma'}^{\phantom \dagger}(0) \rangle \nonumber].
\end{eqnarray}
Here, $ph$ refers to the particle-hole notation\cite{Chi_pp}, 
$\sigma$ and $\sigma'$ denote the spin directions of the
impurity electrons, $T_{\tau}$ is the time ordering operator and $\nu_n$, $\nu_{n'}$ and 
$\Omega_n$ represent two fermionic and one bosonic Matsubara frequency, respectively.  
$\chi_{ph,\sigma \sigma'}^{\nu_n \nu_{n'} \Omega_n}$ can be calculated using
an impurity solver, as described in the following section. 
We recall, that, in the case of SU(2) symmetry, 
the Bethe-Salpeter equation can be diagonalized in the spin sector 
defining the usual charge/spin channels. For this work, the \emph{charge} channel
[$\chi_c^{\nu_n \nu_{n'} \Omega_n} = \chi_{ph, \uparrow \uparrow}^{\nu_n \nu_{n'} \Omega_n} + 
\chi_{ph, \uparrow \downarrow}^{\nu_n \nu_{n'}\Omega_n}$] is of particular interest. 

Note that $\Omega_n$ will be set to zero throughout this work, and is therefore omitted hereinafter.
This is done to perform comparisons of
the results presented here to results of the recent 
literature\cite{Schaefer2013, Schaefer2016}, 
but also because the irreducible vertex divergences appear, systematically, at
lower interaction values for $\Omega_n=0$, compared to cases for $\Omega_n \neq 0$.

\begin{figure}[t]
{{\resizebox{8.0cm}{!}{\includegraphics {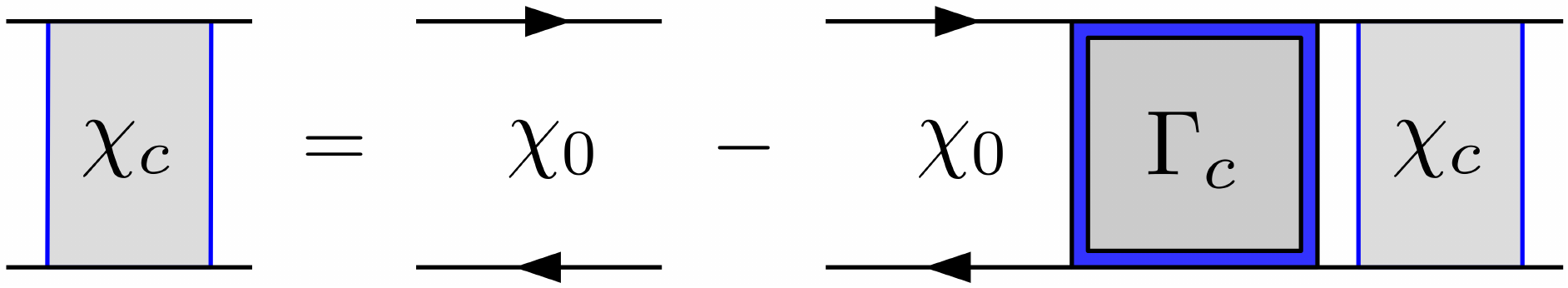}}}
\caption{Schematic representation of the Bethe-Salpeter equation in the charge channel (see text).} 
\label{fig:bs_schematic}
}
\end{figure}

The Bethe-Salpeter equation in the charge channel reads as:

\begin{equation}
\label{equ:bs_equ}
\chi_{c}^{\nu_n \nu_{n'}} = \chi_{ph,0}^{\nu_n \nu_{n'}} - \frac{1}{\beta^2} \sum_{{\nu}_{n_1} {\nu}_{n_2}} 
\chi_{ph,0}^{\nu_n {\nu}_{n_1}} \Gamma_c^{{\nu}_{n_1} {\nu}_{n_2}}  \chi_{c}^{{\nu}_{n_2} \nu_{n'}}
\end{equation}
Here $\Gamma_c^{\nu_n \nu_{n'}}$ is the irreducible vertex function in the charge channel, 
the bare susceptibility is given by 
$\chi_{ph,0}^{\nu_n \nu_{n'} \Omega_n} = -\beta G(\nu_n) G(\Omega_n+\nu_n)
\delta_{\nu_n\nu_{n'}}$. In Fig.\,\ref{fig:bs_schematic} a schematic representation of the 
Bethe-Salpeter equation is given, from which it can be seen, that it represents a two-particle 
analogue to the Dyson equation. 

Inverting Eq.\,(\ref{equ:bs_equ}) and considering $\bm{\Gamma}_{c}$, $\bm{\chi}_{ph, 0}$ 
and $\bm{\chi}_c$ as matrices of the 
fermionic Matsubara
frequencies ($\nu_n,\nu_{n'}$) leads to

\begin{equation}
\label{equ:inv_bs_equ}
\bm{\Gamma}_{c} = \beta^2 \Big( [ \bm{\chi}_c ]^{-1} - [ \bm{\chi}_{ph,0} ]^{-1}   \Big) .
\end{equation}
It is obvious, hence, that all divergences of $\bm{\Gamma}_{c}$ must correspond to a 
singular $\bm{\chi}_c$-matrix\cite{Schaefer2016} 
(typically no divergence is expected in $[ \bm{\chi}_{ph,0} ]^{-1}$). 
In fact, analyzing the matrix in its spectral representation, 
i.e., the basis of its eigenvectors,

\begin{equation}
\label{equ:spectral}
[\bm{\chi}_c ]^{-1}_{\nu_n \nu_{n'}} = \sum_i V^c_{i} (i\nu_{n'})^{\ast}
(\lambda_i)^{-1} V^c_i(i\nu_n) \quad ,
\end{equation}
leads to the 
one-to-one correspondence of an irreducible vertex divergence to a vanishing eigenvalue 
($\lambda_{i=\alpha}\rightarrow 0$) 
of the matrix $\bm{\chi}_c$ in the fermionic frequencies $\nu_n,\nu_{n'}$. 

In particular, for a parameter set ($U, T$) close to a divergence, the
corresponding eigenvalue will be vanishingly small ($\lambda_{i=\alpha}\approx 0$), 
leading to a simplified expression for $\bm{\Gamma}_{c}$:

\begin{equation}
\label{eq:divapprox}
\Gamma_c^{\nu_n\nu_{n'}} \sim \beta^2 {V^{c}_{\alpha}}(i\nu_n)^{\ast} \lambda_\alpha^{-1}
 V^c_{\alpha}(i\nu_{n'}).
\end{equation} 

One sees immediately how in the proximity of a divergence the full frequency structure 
of $\bm{\Gamma}_{c}$, i.e., its dependence on the fermionic  Matsubara frequencies $\nu_n$ and $\nu_{n'}$, is determined\cite{Schaefer2016}
by the non-zero components of the eigenvector $V^c_{\alpha}(i\nu_n)$ associated to the vanishing eigenvalue 
$\lambda_{\alpha}$. 
This leads to a distinction of two classes of irreducible vertex divergences,
a \emph{global} one with an eigenvector $V^c_{\alpha}(i\nu_n) \neq 0 \, \forall \, \nu_n$ 
and a \emph{local} one, 
where only for a finite 
subset of frequencies 
$V^c_{\alpha}(i\nu_n) \neq 0$ holds. 

\begin{figure}[t!]
{{\resizebox{9.2cm}{!}{\includegraphics {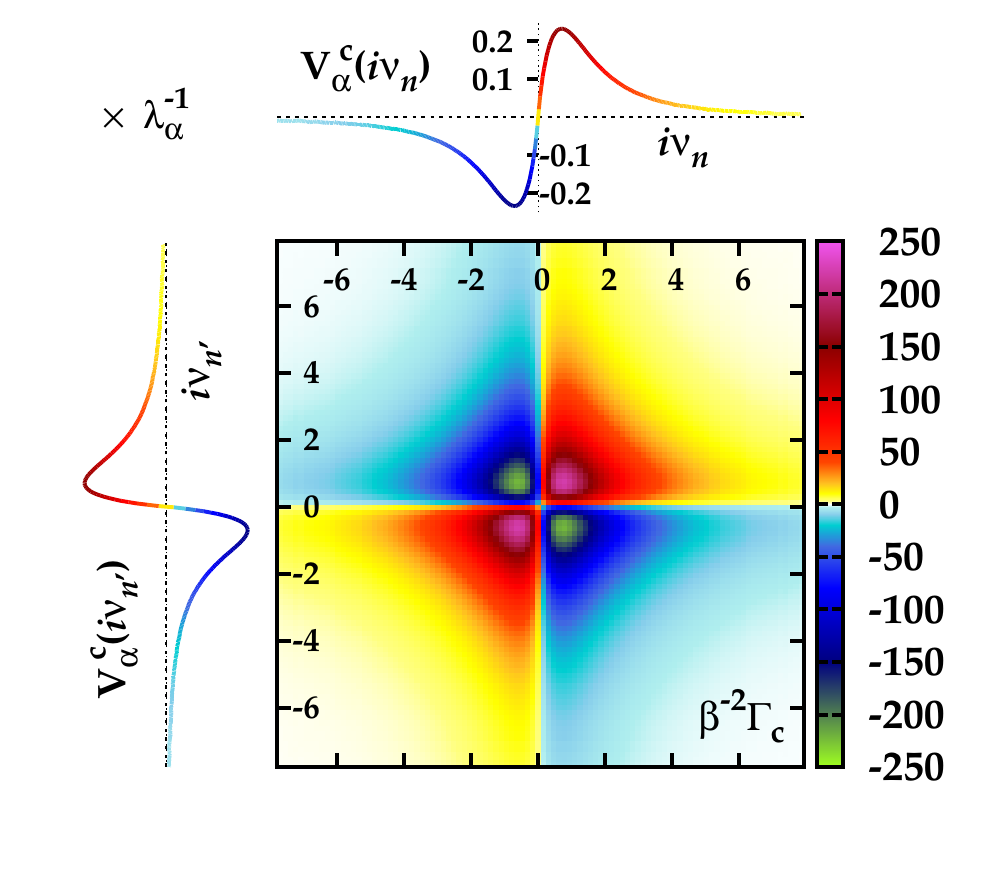}}}
\centering
\caption{\small{Visualization of the relation between the singular eigenvalues $\lambda_\alpha$  
and eigenvectors $V^c_\alpha(i\nu_n)$ of $\bm{\chi}_c$ and the full frequency structure of 
$\bm{\Gamma}_{c}$  in proximity of a divergence (the specific calculation, shown in the main panel as an example, has been performed for an AIM with $U=3.321444$ and $\beta=40$, yielding $\lambda_\alpha =0.00025$). 
The color of the singular eigenvectors (not related to the color scale of the main plot) highlights the connections to the sign/intensity structure of $\bm{\Gamma}_{c}$ as a function of $\nu_n$, $\nu_{n'}$, defined by  Eq.\,(\ref{eq:divapprox}).  The values 
of $\bm{\Gamma}_{c}$ are rescaled by $1/\beta^2$ in the main panel for a better readability.}}
\label{fig:vertex_schematic}
}
\end{figure}

The interplay of eigenvectors, eigenvalues and $\bm{\Gamma}_{c}$ is further discussed in 
Sec.\,\ref{sec:singEV_class}. 
Already at this stage, however, we illustrate how the direct relation of Eq.\,(\ref{eq:divapprox}) is 
actually realized in the proximity of a divergence: In Fig.\,\ref{fig:vertex_schematic} we show a 
pertinent example of the vertex function  computed  for a parameter set very close to a divergence,   
where the lowest eigenvalue $\lambda_\alpha$ of $\bm{\chi}_c$ is $O(10^{-4})$. In this figure, the full 
(fermionic) Matsubara frequency dependence ($\nu_n, \nu_{n'}$) of $\bm{\Gamma}_{c}$ is plotted (main panel) 
together with the eigenvector $V^c_\alpha(i\nu_n)$ 
(both on the left and on top of the main panel) associated to 
the smallest, almost vanishing, eigenvalue $\lambda_\alpha$. 
It can be easily noticed how in the proximity of a vertex divergence, the frequency structure of 
$\bm{\Gamma}_{c}$, including 
the location of the maxima/minima and its signs, is completely controlled by the corresponding frequency dependence of the singular eigenvector $V^c_\alpha(i\nu_n)$. The latter encodes, thus, all the essential information about the divergence itself, and will be used in the following for analyzing the evolution of the frequency structure of the vertex function in the proximity of different divergences.

Note that in Sec.\,III also results for the divergences in the particle-particle up-down channel
are shown [$\chi_{pp, \uparrow \downarrow}^{\nu_n \nu_{n'}}$]. 
For this channel the same general consideration made here holds, the corresponding Bethe-Salpeter 
equation can be found in Appendix B of Ref.\,[\onlinecite{Rohringer2012}] and reads in 
particle-particle notation

\begin{eqnarray}
\chi_{pp, \uparrow \downarrow}^{\nu_n (-\nu_{n'})} =  
-\frac{1}{\beta^2}\sum_{{\nu}_{n_1} {\nu}_{n_2}} 
 \Big(\chi_{pp, 0}^{\nu_n {\nu}_{n_1}} - 
\chi_{pp, \uparrow \downarrow}^{\nu_n (- {\nu}_{n_1}) } \Big)
\Gamma_{pp, \uparrow \downarrow}^{{\nu}_{n_1} {\nu}_{n_2} } 
\chi_{pp, 0}^{{\nu}_{n_2} \nu_{n'}} \nonumber 
\end{eqnarray}

Let us also briefly comment, at the end of this section, on the degree of two-particle irreducibility of the vertex considered.
While the vertex we obtain by the inversion of the Bethe-Salpeter equation in a given channel (e.g. the charge channel) is, per 
construction, 2PI only in that specific channel, its divergences  correspond\cite{Schaefer2013,Schaefer2016} precisely to the divergences of the fully 2PI vertex 
function. In this respect, we recall that the vertex divergences found here are {\sl not} associated to any thermodynamic phase transition, and  never appear in the full two-particle scattering amplitude ($F$). Hence, due to the algebraic structure of the parquet equation\cite{Bickers,Rohringer2012}, if one of such a divergence occurs, e.g. in
 $\bm{\Gamma}_{c}$, it has to be compensated by an analogous divergence of the fully 2PI vertex function, in order to preserve the finiteness of $F$  (for more details see [\onlinecite{Schaefer2016,Rohringer_REV}]). This has been explicitly verified also for the vertex divergences discussed in the following sections.
 
We note that the choice of studying the divergences in $\bm{\Gamma}_{c}$, instead of considering the equivalent ones in the fully 2PI vertex, is also suggested by the more direct connection of  $\bm{\Gamma}_{c}$ to the LW functional (of which $\bm{\Gamma}_{c}$ represents the second functional derivative) and, hence, to the previously mentioned multivaluedness issues\cite{Kozik2015,Gunnarsson2017, Vucevia2018}. 

\subsection{Calculations in CT-QMC}
\label{sec:calculation_ctqmc}

We solve the AIM using a continuous-time quantum Monte Carlo (CT-QMC) impurity solver
in the hybridization expansion\cite{Werner2006, Werner2006b}.
The algorithm is based on a stochastic Monte Carlo sampling of the infinite series expansion of the
partition function in terms of the hybridization. 

From the stochastic series expansion of the partition function one can construct estimators for the 
one-particle and two-particle Green's function and, thus, the generalized susceptibility defined in 
Eq.\,(\ref{equ:form_gen_chi}). Extracting the irreducible vertex function from the one- and 
two-particle Green's functions, by inverting the corresponding Bethe-Salpeter equation, 
is a post-processing step to the Monte Carlo measurement, as is the calculation of eigenvalues
and eigenvectors of the generalized susceptibility.

Further, we recall\cite{Schaefer2016} that for an 
easier numerical identification of the singular eigenvalues and eigenvectors it is convenient 
to diagonalize $(\chi_{c}/\chi_{ph,0})^{\nu_n \nu_{n'}}$ instead of $\chi_{c}^{\nu_n \nu_{n'}}$.
This way it is straightforward to distinguish the vanishing 
eigenvalues of $\chi_{c}^{\nu_n \nu_{n'}}$ from the trivial high-frequency 
eigenvalues ($\propto 1/\nu_n^2$) and the corresponding eigenvectors. 
The results are not influenced by this procedure, because the vanishing of 
an eigenvalue of $\chi_{c}^{\nu_n \nu_{n'}}$ corresponds to the one of
$(\chi_{c}/\chi_{ph,0})^{\nu_n \nu_{n'}}$ 
($\chi_{ph,0}^{\nu_n \nu_{n'}}$ is not singular). Hence, the 
specific interaction value for a given temperature where a vertex
divergence occurs, i.e. $\widetilde{U}$, is 
identical. Further, for all cases considered in this work, 
the numerical difference between the corresponding eigenvectors  
was found to be negligible.
Hence, in the rest of the paper we will consider identical -- for all practical 
purposes -- the singular eigenvectors of $(\chi_{c}/\chi_{ph,0})^{\nu_n \nu_{n'}}$ and
$\chi_{c}^{\nu_n \nu_{n'}}$.  
The details of the procedure for determining $\widetilde{U}$ for a given temperature, 
is described in the Appendix C.

For the specific CT-QMC calculations, of the one- and two-particle Green's function needed in this 
work, we have employed the w2dynamics software\cite{Parragh2012}. 
The vertex functions generated by w2dynamics were previously tested against 
other established codes\cite{Ribic2017}. Additionally, we have benchmarked the 
reliability of the impurity solver in computing the vertex divergences of the AIM, against 
exact diagonalization (ED) results in an intermediate $T$-region,
where the discretization of the electronic bath affects the ED procedure only moderately.

For the low temperature calculations, we have quantified the reliability 
of our results using a Jackknife error analysis\cite{Jackknife}, 
which is described in the Appendix C.

\section{Numerical Results}

\subsection{The T--U diagram}
\label{sec:tudiagram}

\begin{figure*}[t!]
{{\resizebox{8.9cm}{!}{\includegraphics {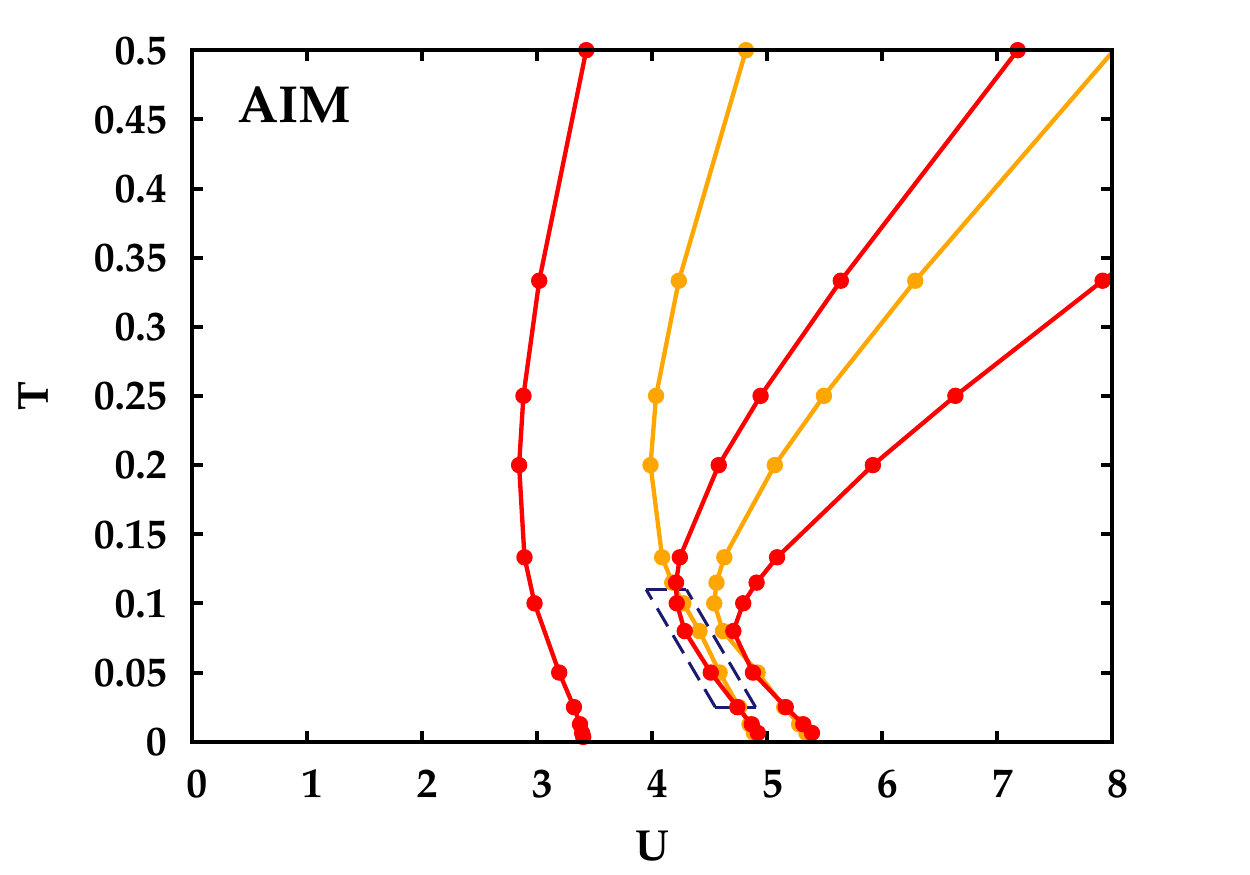}}}}
{{\resizebox{8.9cm}{!}{\includegraphics {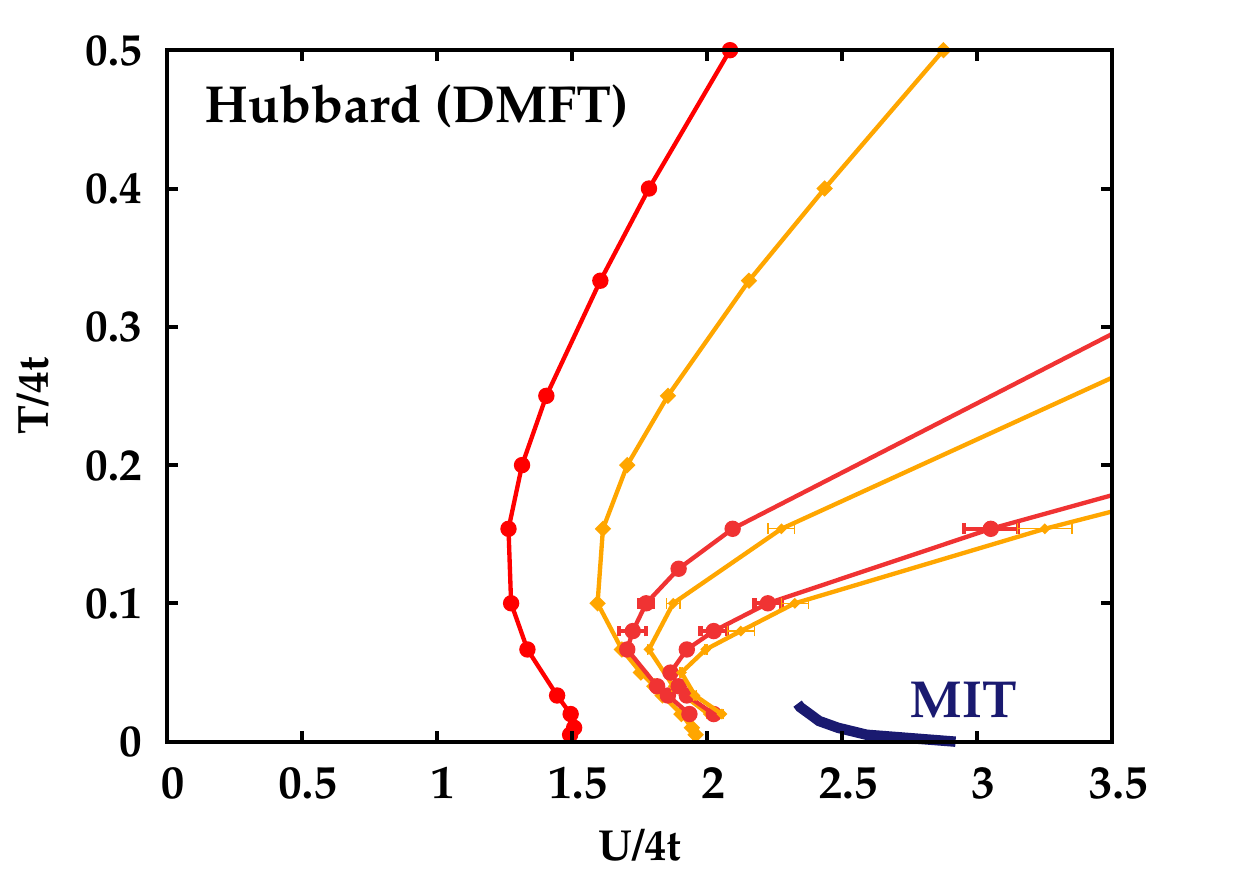}}}
\caption{Left panel: $T-U$ diagram of the AIM at half-filling, showing the first divergence lines 
along which the irreducible 
vertex functions diverge. For red lines this divergence takes place in the charge channel,  
$\Gamma_c^{\nu_n\nu_{n'}(\Omega_n=0)}$, along the orange lines simultaneous divergences in the 
charge and the particle-particle 
up-down channel, $\Gamma_c^{\nu_n\nu_{n'}(\Omega_n=0)}$ and
$\Gamma_{pp, \uparrow \downarrow}^{\nu_n\nu_{n'}(\Omega_n=0)}$, are observed. 
The dashed blue box marks the parameter region where the "atomic" ordering of divergence lines
is violated (see text).  
Right panel: 
Divergence lines of the half-filled unfrustrated Hubbard model (square lattice dispersion with 
$4t=1$), solved with DMFT. The lines are plotted 
with the same color code, the blue solid line represents the Mott-Hubbard MIT\cite{Bluemer_PhD}. 
Re-adapted from Ref.\,[\onlinecite{Schaefer2016}]. } 
\label{fig:phase_diag_all}
}
\end{figure*} 

We start to illustrate our numerical results by reporting in the $T$--$U$ diagram of the AIM 
(Fig.\,\ref{fig:phase_diag_all} left panel) the first (five) 
lines along which the two-particle irreducible vertex diverges. 
These correspond to the interaction values
$\widetilde{U}$ at given temperatures $T$, where an eigenvalue of the generalized susceptibility 
(charge or particle-particle up-down channel) vanishes, see Eq.\,(\ref{equ:inv_bs_equ}) and 
Eq.\,(\ref{equ:spectral}). Specifically, the red lines mark irreducible 
vertex divergences taking place in the charge channel only, while 
orange lines represent divergences taking place in the 
charge and the particle-particle up-down channel simultaneously. 
  
Even from the first look at the data, the overall behavior of the divergence lines of the AIM appears  
qualitatively \emph{very similar} to the one of the Hubbard model case\cite{Schaefer2013,Schaefer2016}, 
reproduced in the right panel of Fig.\,\ref{fig:phase_diag_all} 

In particular, the similarity in the 
high-temperature/large interaction area of both $T$--$U$ diagrams is not fully unexpected. 
In fact, here the divergence lines of 
both models display a rather linear 
behavior, which is consistent with the insights obtained from the results 
of the Hubbard atom case\cite{Schaefer2016}.
The residual deviations can be ascribed to 
the fact that the atomic limit condition,
i.e., $U$ and $T$ larger than all other energy scales, is not fully complied.
In the case of the AIM, only for larger interactions than those shown in the left panel of 
Fig.\,\ref{fig:phase_diag_all}
($U \geq D=10$), we recover a purely linear behavior as well as the 
connection between the position of the divergence line and the inflection point 
of $\mathrm{Im}\, G(i\nu_n)$, as expected for the atomic limit (for a more detailed analysis, 
see Appendix A).

At intermediate temperatures, the divergence lines show a progressively stronger non-linear behavior, 
starting to bend rightwards.  
Lowering the temperature further, one reaches the correlated metallic regime.  
Remarkably, in spite of the differences in the ground states of the two models
(there is no MIT in the AIM), 
even there the results of the AIM and the Hubbard
model remain qualitatively \emph{very similar}. For both models the lines show a "re-entrance", i.e.
a bending towards higher interaction values, as if the low-temperature intermediate interaction regime
were "protected" against the non-perturbative mechanism originating the irreducible vertex 
divergences. Particularly remarkable, however, is that 
finite $\widetilde{U}$ values at $T=0$ are observed in both cases, for the AIM the 
low-temperature behavior of the first line is investigated in detail in Sec.\,\ref{sec:lowT_red1}.

\begin{figure}[tb]
{{\resizebox{9.1cm}{!}{\includegraphics {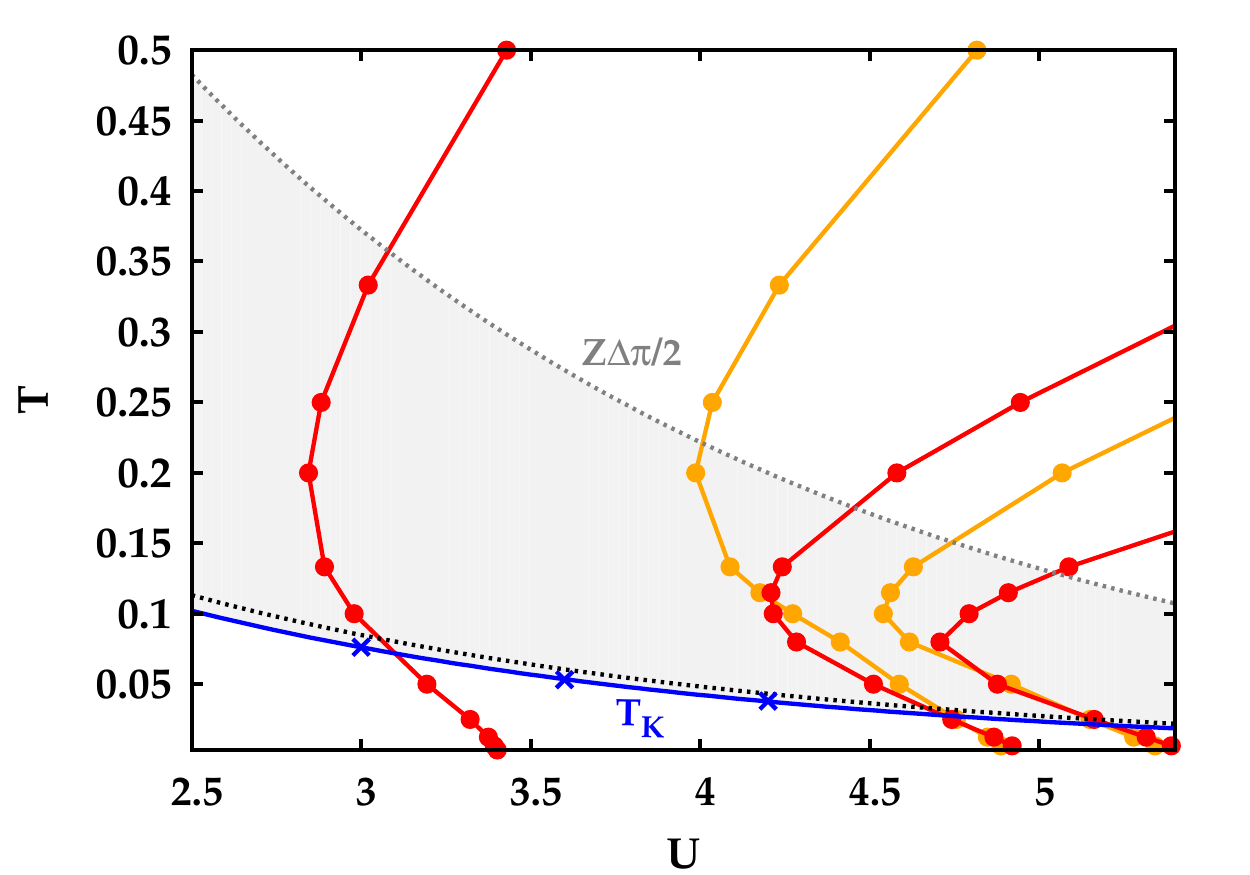}}}
\caption{A zoom of the $T$--$U$ diagram of the AIM (left panel Fig.\,\ref{fig:phase_diag_all}) 
at half-filling is shown. 
The blue-solid line marks the Kondo temperature ($T_K$), estimated from the
rescaling of our numerical data for the magnetic susceptibility to the 
universal function given in Ref.\,[\onlinecite{Krishna}]. The 
black-dotted line represents an estimate for $T_K$ obtained from an analytic 
expression\cite{Hewson-precise} valid in
the limit $D \gg U,T$. 
An additional scale related to the Kondo screening, the half-bandwidth of 
the $T \rightarrow 0$ Kondo peak 
($\frac{\pi}{2}Z\Delta$) \cite{Hewson} is marked with a gray-dotted line, and is
roughly five times larger than $T_K$. The light-gray shaded area can be regarded, thus, 
as the parameter region 
where the effects of the Kondo screening become visible.} 
\label{fig:kondo_phasediag}
}
\end{figure}

In the framework of the 
overall similarity discussed above, a specific difference can 
be seen, however. This is 
highlighted by the dashed blue box in the left panel of Fig.\,\ref{fig:phase_diag_all} (see 
Fig.\,\ref{fig:kondo_phasediag} for a zoom): 
At intermediate temperatures, 
the second and third divergence line in the $T$--$U$ diagram of the AIM cross, 
breaking the typical line-order found in all cases 
analyzed so far in the 
literature\cite{Schaefer2013, Schaefer2016, Gunnarsson2016, Ribic2016, Gunnarsson2017}  
(i.e., always an orange line 
after a red one, before the next red line). 
The two divergence lines, however, cross again 
at lower temperatures, restoring 
the typical line-order. We also observe, that even the fourth and fifth line 
show such a peculiar crossing, though, to a much smaller extent.
To verify the reliability of this observation several tests were performed using 
exact diagonalization (ED) calculations of the generalized susceptibility\cite{ED_Check}. As it turns out, 
our ED analysis (not shown) 
has confirmed, within the numerical accuracy, the occurrence of such a line crossing.

Although somewhat unexpected and unobserved in preceding studies, 
the crossing of divergence lines is, however, not in conflict with the most
recent theoretical progress made in the analysis of vertex divergences (see 
Ref.\,[\onlinecite{Gunnarsson2017}]).
In that work, it has been demonstrated that vertex divergences 
of both kinds are originated by the crossings between different branches of the 
Luttinger-Ward functional (LWF) of the self-energy. While in the cases considered 
hitherto\cite{Gunnarsson2017, Vucevia2018} crossings of at most two branches have been reported, 
it can be logically inferred, that due to the existence of infinite unphysical branches, for 
other choices of models/parameters (such as in our AIM) crossings among three (or more) branches 
of the LWF (of which, of course, only one is physical) occurs\cite{Cross}. The intersections 
of two divergence lines observed in our calculations then suggest that this indeed happens for the 
AIM considered here. It remains to understand, however, why such a situation is, apparently, not 
realized in the correlated metallic regime of the Hubbard model solved by DMFT.   

Finally, as for the theoretical understanding of the low-$T$ regime of the AIM, it is 
important to estimate the Kondo scale $T_K$ and its 
possible connection to the properties of the irreducible vertex divergences.
In Fig.\,\ref{fig:kondo_phasediag} a zoom of the $T$--$U$ diagram of the AIM shown in 
Fig.\,\ref{fig:phase_diag_all} is presented together with several estimates 
for the Kondo temperature $T_K$.
In particular, the black dotted line represents an analytic estimate valid in the $D\gg U,T$ parameter 
regime\cite{Hewson-precise} 
($T_K=0.4107 U (\frac{\Delta}{2U})^{1/2} e^{-\pi U/8\Delta + \pi \Delta /2U}$, 
where in our AIM: $\Delta = \pi \rho_0 V^2 = \pi/5$), while the blue line
is determined through the universal scaling of the numerical susceptibility
data\cite{Krishna} (see Appendix B).
We note that the two procedures yield extremely close estimates of $T_K$.
The Kondo temperature marks however not a phase transition but a smooth crossover. Indeed,
the screening processes associated with it become active 
already at temperatures larger than $T_K$.
For instance, we see that the temperature below which the effects of the Kondo resonance 
become visible in the spectrum is $T \lesssim Z \Delta \frac{\pi}{2}$, the half-bandwidth of the central 
peak\cite{Hewson}.
We choose this scale to define the upper border of the corresponding crossover regime
(shaded gray area in the $T$--$U$ diagram of Fig.\,\ref{fig:kondo_phasediag}). 
It is quite visible, how the bending of the divergence lines is essentially occurring in this 
parameter region.

\subsection{Classification of the singular eigenvectors}
\label{sec:singEV_class}

\begin{figure*}[t!]
{{\resizebox{8.0cm}{!}{\includegraphics {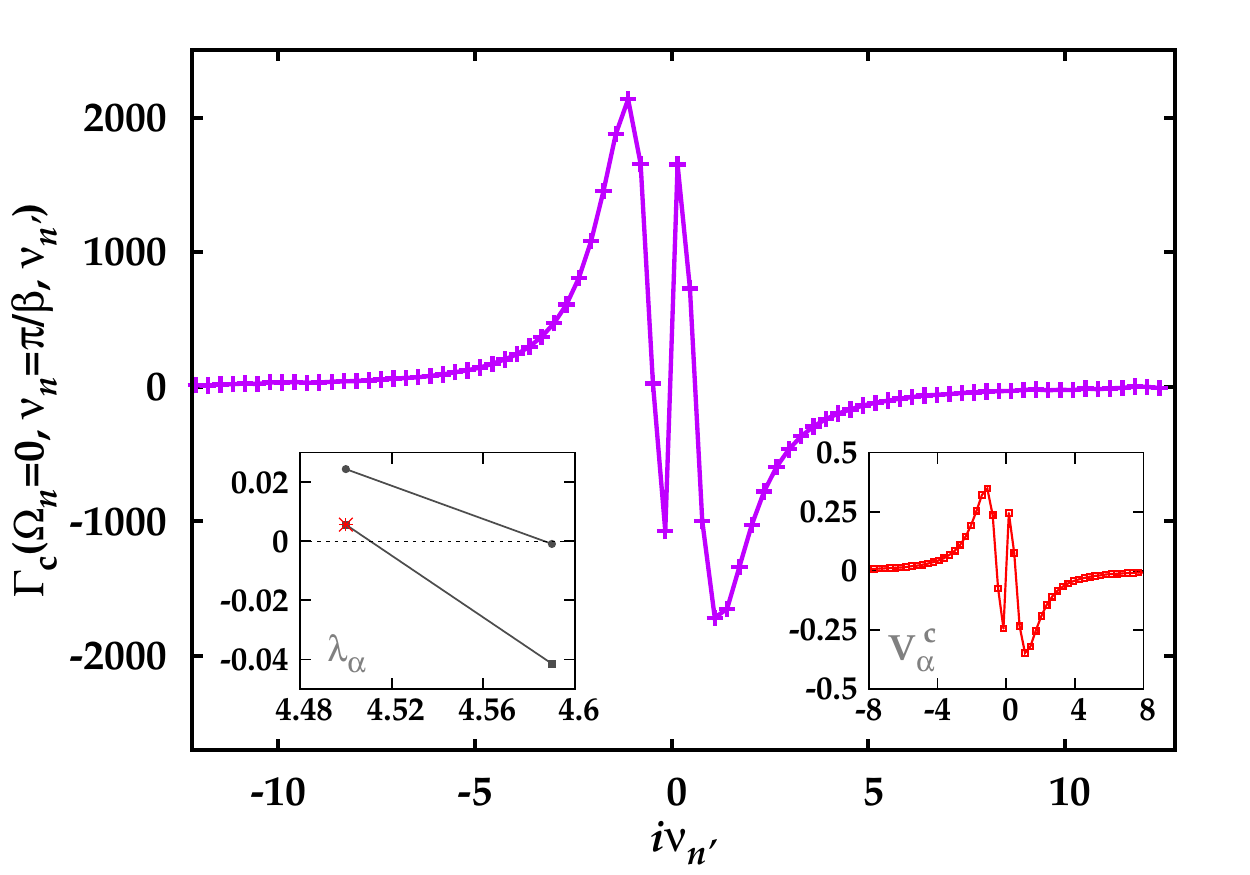}}}}
{{\resizebox{8.0cm}{!}{\includegraphics {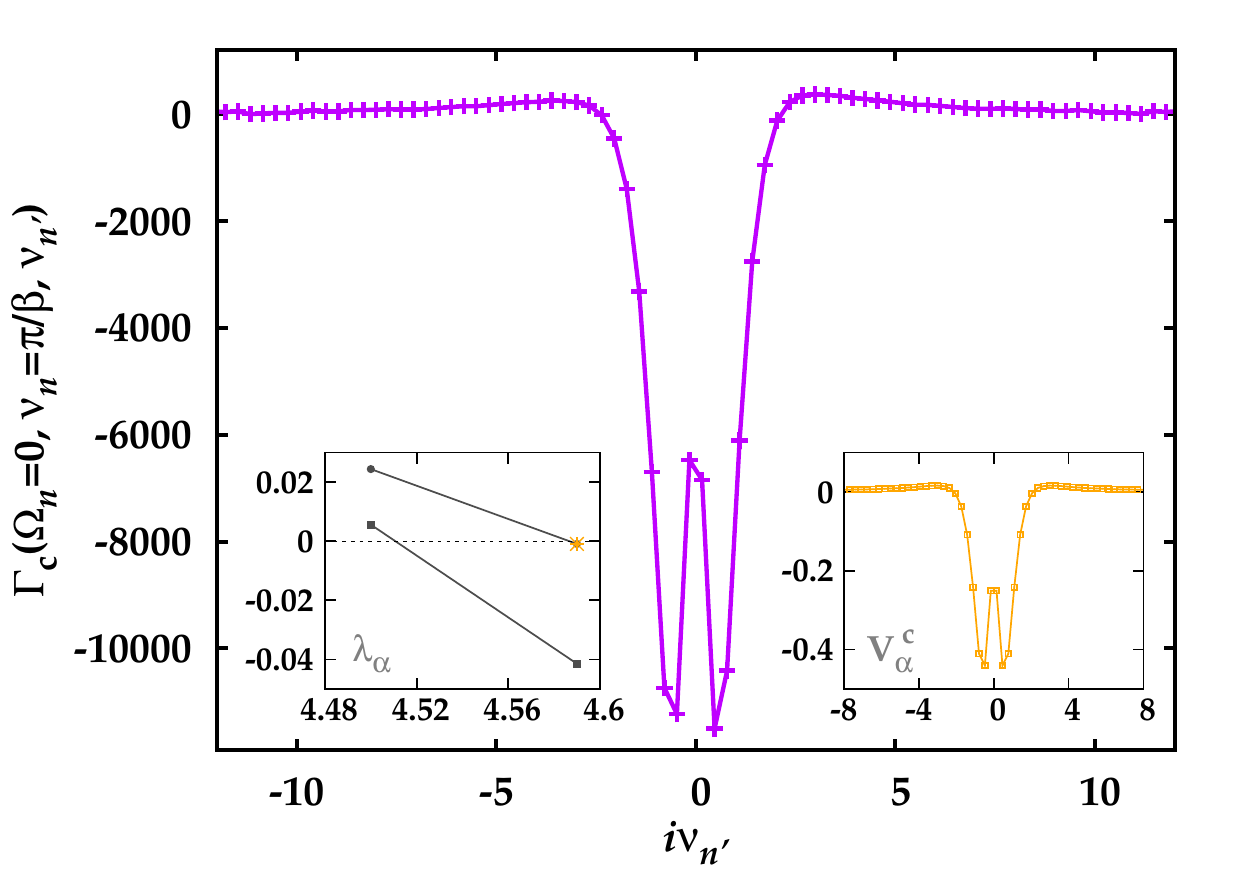}}} 
\caption{Cuts of the irreducible vertex function in the charge channel, 
$\Gamma_c^{\nu_n\nu_{n'}(\Omega_n=0)}$, for the first Matsubara frequency, $\nu_1=\pi/\beta$ at $T=0.05$
and two $U$ values are shown. 
In the left insets the lowest two eigenvalues 
of $(\chi_c/\chi_{ph,0})^{\nu_n\nu_{n'}(\Omega_n=0)}$ are reported vs. $U$, 
in the right insets the 
eigenvector $V_{\alpha}^c(i\nu_n)$ corresponding 
to the lowest eigenvalue $\lambda_{\alpha}$ 
is shown as a function of $i\nu_n$. 
Left panel: At $U=4.5$ the lowest eigenvalue is corresponding to 
the second red divergence line (red dot), hence $V_{\alpha}^c(i\nu_n)$ 
is antisymmetric. Right panel: For 
$U=4.59$ the eigenvalue of the first orange line is the smallest (orange dot), 
$V_{\alpha}^c(i\nu_n)$ is symmetric.} 
\label{fig:ev_gamma_symm}
}
\end{figure*}

In order to make our study of the vertex divergences in the AIM more quantitative, we 
proceed with the analysis of the singular eigenvectors in the charge channel, associated to a 
vanishing eigenvalue of $\chi_c^{\nu_n \nu_{n'}}$ (see Eq.\,(\ref{equ:spectral})).
In fact, as mentioned in Sec.\,\ref{sec:formalism}, their frequency structure controls 
the frequency dependence of $\Gamma_c$ in the proximity of -- and especially at -- a vertex divergence. 
We note, that for the orange divergence lines, where $\Gamma_c$ and $\Gamma_{pp, \uparrow \downarrow}$ 
diverge simultaneously, the frequency structure 
of the singular eigenvectors $V^c_{\alpha}(i\nu_n)$ and 
$V^{pp, \uparrow \downarrow}_{\alpha} (i\nu_n)$ is found to be identical, which is why 
$V^{pp, \uparrow \downarrow}_{\alpha} (i\nu_n)$ will not be shown in the following.

\begin{figure}[tb!]
{{\resizebox{8.1cm}{!}{\includegraphics {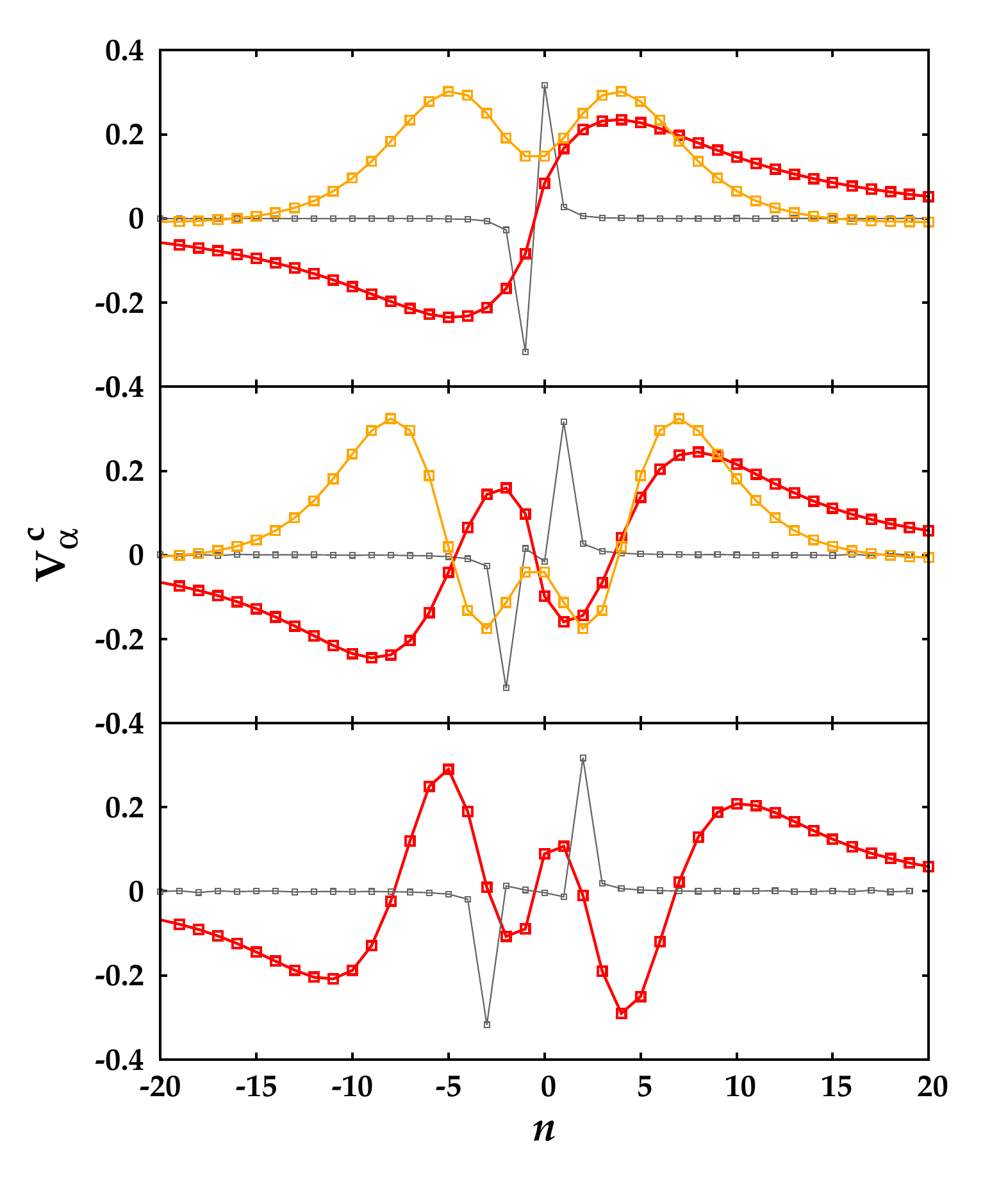}}}
\caption{Singular eigenvectors 
of $(\chi_c/\chi_{ph,0})^{\nu_n\nu_{n'}(\Omega_n=0)}$ (numerically equivalent to the ones 
of $\chi_c^{\nu_n\nu_{n'}(\Omega_n=0)}$), corresponding 
to the five divergence lines (left panel 
Fig.\,\ref{fig:phase_diag_all}), shown as a 
function of Matsubara index $n$ for the 
temperature $T=0.025$. 
In gray the eigenvectors of the red divergence lines for a higher temperature, $T=0.5$, properly
rescaled, are plotted, indicating the broadening of $V^c_{\alpha}(i\nu_n)$ for
lower temperatures. Top panel: 
Eigenvectors of the first red divergence line (red, antisymmetric) and the 
first orange divergence line (orange, symmetric). Middle (Bottom) panel: Same as top panel, but for 
the second (third) red and orange (red) divergence line.} 
\label{fig:ev_red_orange_loc_sym}
}
\end{figure}

Before showing our numerical results, we discuss some general properties, applicable to a
particle-hole and time-reversal symmetric case, like our AIM. 
In particular, the particle-hole symmetry implies that $\chi_c^{\nu_n \nu_{n'}}$,
considered as a matrix of the two fermionic Matsubara frequencies, 
is a \emph{centrosymmetric} matrix, i.e., it is invariant under a 
$\nu_n\rightarrow -\nu_n$, $\nu_{n'}\rightarrow -\nu_{n'}$ 
transformation\cite{Rohringer2012, Rohringer_Diss}. 
A centrosymmetric matrix in Matsubara frequency space has the property that 
its (non-degenerate) eigenvectors are either symmetric or antisymmetric\cite{Centrosymm}. 
Indeed, our results show that eigenvectors 
associated to red divergence lines are antisymmetric under the transformation 
$\nu_n \rightarrow -\nu_n$, whereas orange eigenvectors are symmetric, as it can be seen in the right
insets of  
Fig.\,\ref{fig:ev_gamma_symm} and in Fig.\,\ref{fig:ev_red_orange_loc_sym}.
The symmetry of the singular eigenvectors is -- as expected -- well reflected 
in the frequency structure of the irreducible vertex. 
As an illustrative example, a cut of the irreducible vertex function 
in the charge channel, $\Gamma_c^{\nu_n=\pi T, \nu_{n'}, \Omega_n=0}$ for two values 
of the interaction $U$ at the same temperature ($T=0.05$) is shown in Fig.\,\ref{fig:ev_gamma_symm}. 
In fact, in spite of the proximity between the second red and the first orange divergence line
for these parameters, it can be clearly seen 
how the frequency structure of the vertex function is almost perfectly antisymmetric/symmetric
in the case where the lowest eigenvalue corresponds to a red/orange divergence line (left/right panel).

After discussing this general 
feature of the singular eigenvectors, applicable to all particle-hole symmetric 
models hitherto analyzed\cite{Schaefer2016}, we turn to 
their intriguing evolution with decreasing temperature, and start by going back to 
Fig.\,\ref{fig:ev_red_orange_loc_sym}. 
There, eigenvectors 
corresponding to the five divergence lines (three red, two orange) shown in the 
left panel of Fig.\,\ref{fig:phase_diag_all} are compared for 
the same temperature ($T=0.025$). We further plot properly rescaled 
eigenvectors corresponding to the red 
lines at the highest 
temperature employed in the calculations ($T=0.5$) in gray.
The latter show an almost perfect agreement with the atomic limit:  
Eigenvectors, localized in Matsubara frequency space, which have finite 
weight almost only at one frequency [$\nu_n=(2n+1)\pi T$] equal to the 
energy scale $\nu^{\ast}$. For example 
for the first divergence line (top panel)
the gray eigenvector displays its by far largest contribution at the first Matsubara frequency ($n=1$).

This specific property of frequency localization characterizing the singular eigenvectors 
of the red divergence lines (see Sec.\,\ref{sec:formalism}) gets lost, however, 
when reducing the temperature. 
At $T=0.025$ (red and orange eigenvectors) we note that 
their frequency decay is even slower than for the singular eigenvectors 
of the orange lines, which are always associated to "global" divergences, even in the AL\cite{Schaefer2016}.
This means, in turn, that also the divergence of $\Gamma_c$ is 
no longer restricted  to a finite set of 
frequencies. Such a "frequency-broadening" of the red singular eigenvectors at low temperatures
was so far only observed in the DMFT solution of the Hubbard model\cite{Schaefer2016}, and seems to be
associated with the presence of coherent quasiparticle excitations.

\begin{figure*}[t]
{{\resizebox{8.4cm}{!}{\includegraphics {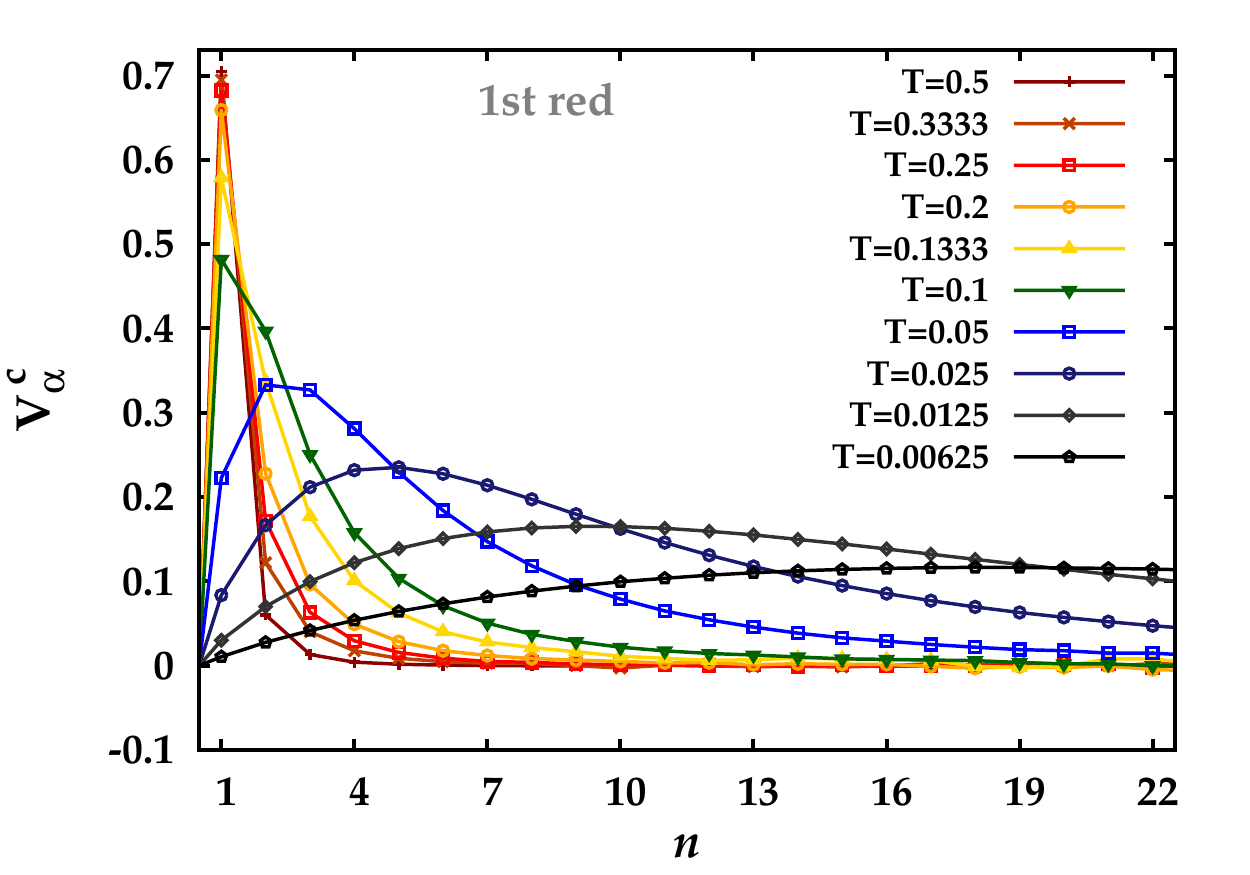}}}}
{{\resizebox{8.4cm}{!}{\includegraphics {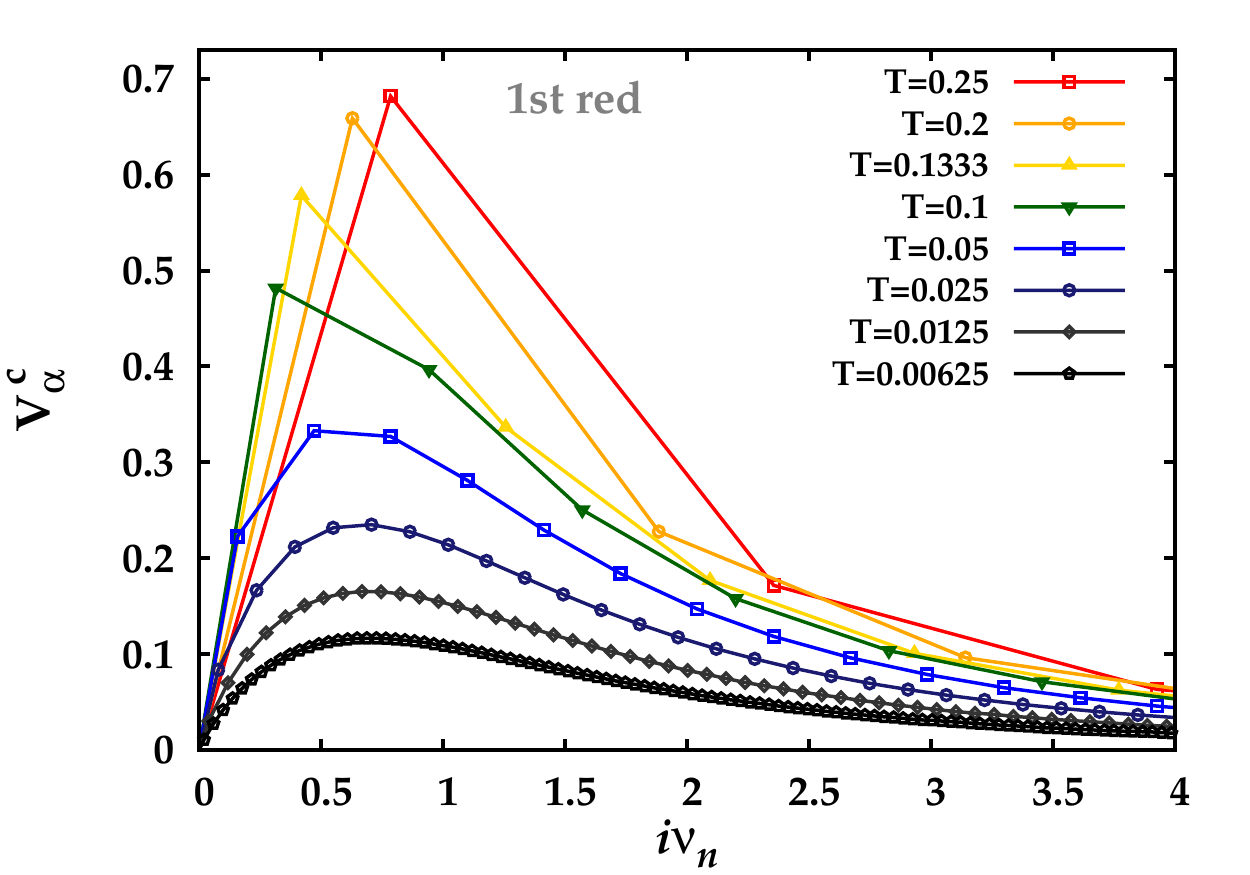}}}}
{{\resizebox{8.4cm}{!}{\includegraphics {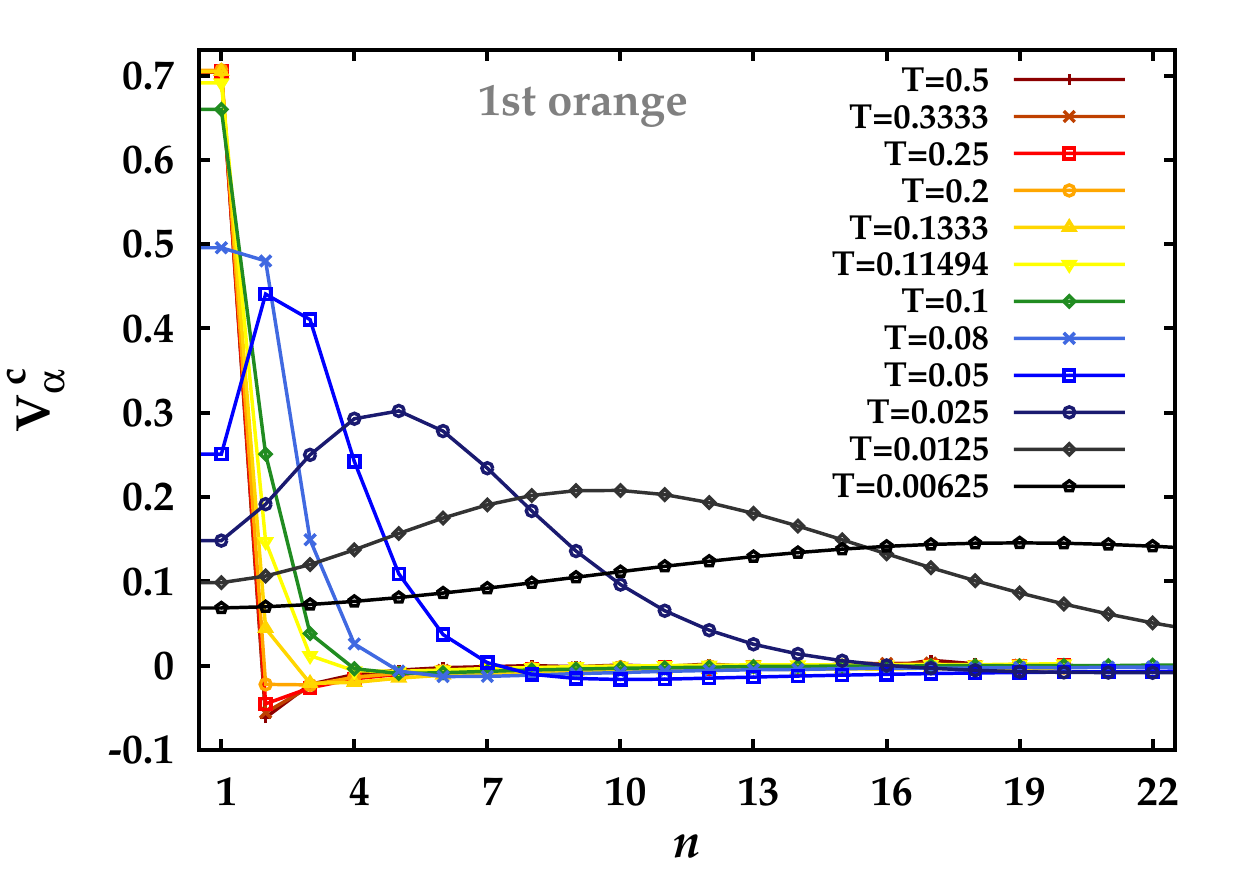}}}}
{{\resizebox{8.4cm}{!}{\includegraphics {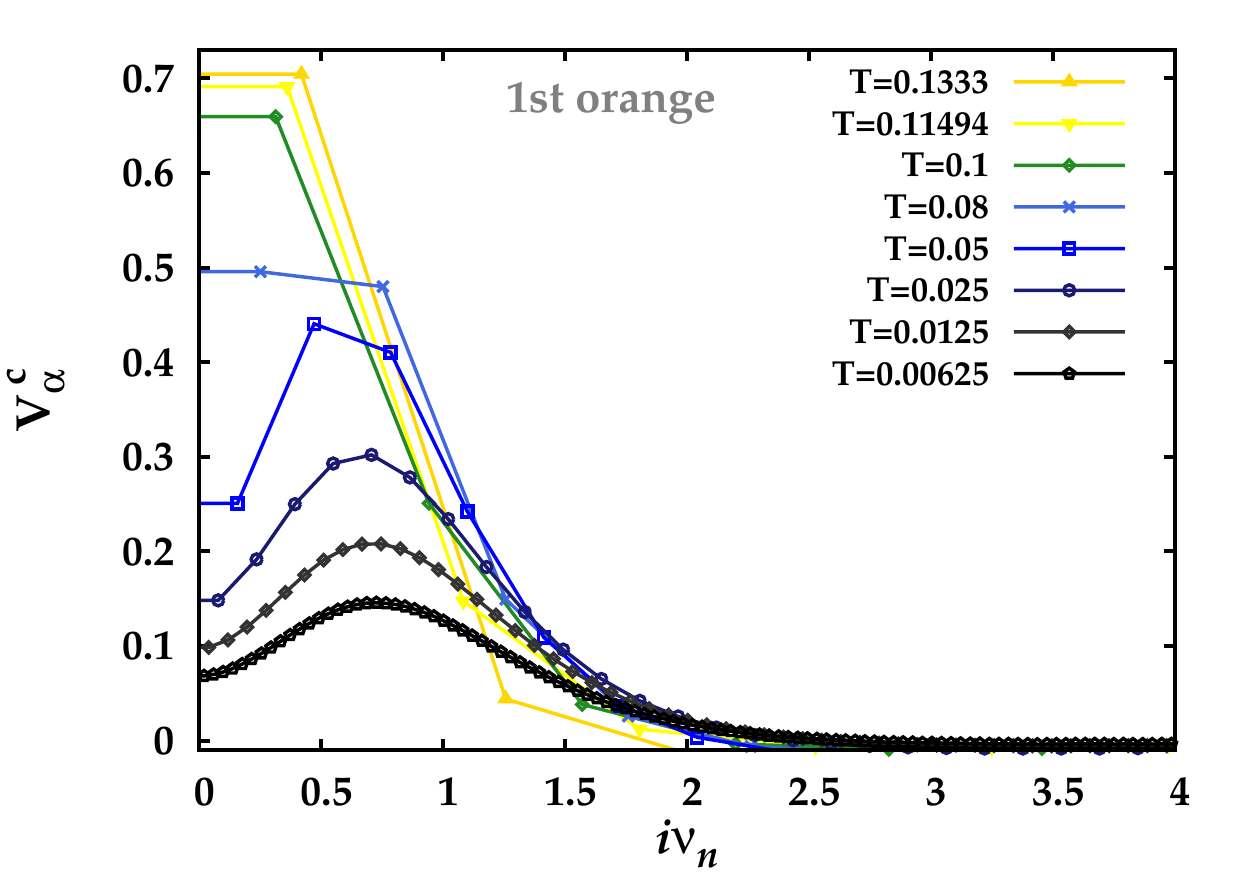}}} 
\caption{Left top panel: Singular eigenvectors $V^c_{\alpha}(i\nu_n)$ 
for several temperatures along the first 
red divergence line, plotted as a function of the Matsubara index $n$. 
Right top panel: Singular eigenvectors of 
the left top panel plotted as a function of Matsubara frequency $i\nu_n$ instead. 
Bottom panels: As top panels,
but showing the data corresponding to the first orange divergence line.} 
\label{fig:T_evolution_ev}
}
\end{figure*} 

This general trend is analyzed in detail in Fig.\,\ref{fig:T_evolution_ev}: In the 
left panels the eigenvectors are plotted in terms of the 
Matsubara index $n$, while in the right panels several $V^c_{\alpha}$ for 
low temperatures are reported as a function of Matsubara frequency $i\nu_n$. It can be easily seen, then, 
that for the eigenvectors corresponding to the first red (upper) and the first orange (lower) divergence 
line 
two regimes are distinguishable: ($i$) for $T \gg T_K$, 
the $V^c_{\alpha}$  
are strongly peaked at a given Matsubara index $n_{\text{max}}$, in perfect agreement with the 
results of the AL. 
($ii$) for $T \lesssim T_K$ the maximum contribution of the eigenvector moves to a higher index with 
decreasing temperature, i.e.\ to the right (Fig.\,\ref{fig:T_evolution_ev} left panels). 
Remarkably, one notices instead that, as a function of Matsubara frequency, 
the maximum contribution of $V^c_{\alpha}(i\nu_n)$
remains localized at a given frequency $i\nu_{n_{\text{max}}}$ in this regime 
(Fig.\,\ref{fig:T_evolution_ev} right panels).  

Finally, it is interesting to analyze in more detail the low-frequency structures
of $V^c_{\alpha}(i\nu_n)$, which can be highlighted by comparing the 
red singular eigenvectors of different lines at the same low temperature, see 
Fig.\,\ref{fig:ev_red_orange_loc_sym}.
In particular, for the eigenvector of the second red divergence line (middle panel of 
Fig.\,\ref{fig:ev_red_orange_loc_sym}) an additional local maximum and  
minimum appear at the lowest frequencies, 
leading to three "nodes" in their frequency components.
In the case of the third red line (bottom panel) $V^c_{\alpha}(i\nu_n)$
has five "nodes". 
Extrapolating the behavior observed for the first three red divergence lines, one 
expects that 
the eigenvector of the $n$-th red divergence line will have $2n-1$ "nodes". 
It is also interesting to note, that for the eigenvectors of the first and second red 
divergence line the respective one or three nodes are also observed in the 
high-$T$ regime (see the gray eigenvectors). 
This, however, no longer holds for the eigenvector of the third line. 

\subsection{Calculations in the low-T regime}
\label{sec:lowT_red1}

\begin{figure}[t]
{{\resizebox{8.0cm}{!}{\includegraphics {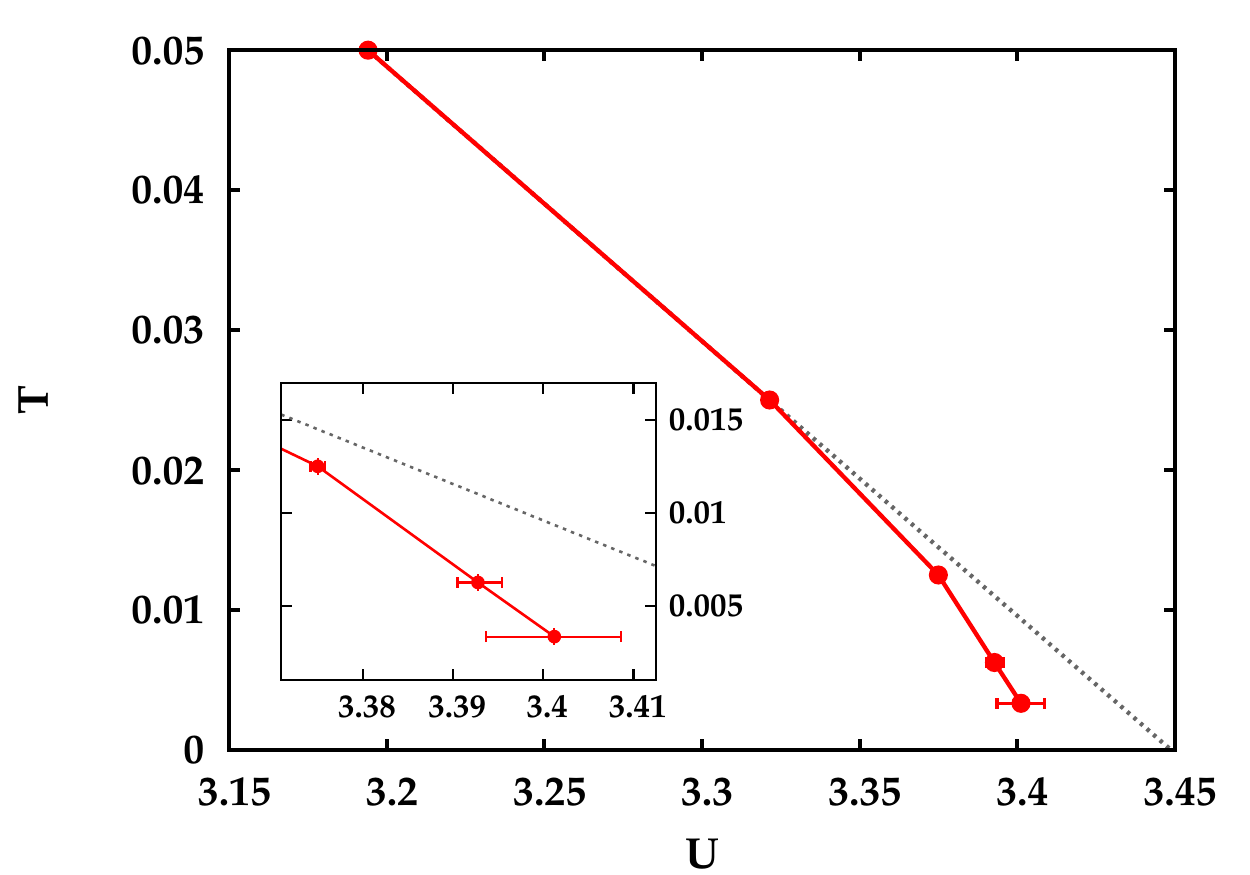}}}
\caption{Zoom on the first red 
divergence line showing the low-$T$ behavior (for $T \ll T_K$) which shows, that within the error bars 
obtained by a Jackknife analysis (see Appendix C) 
the line bends towards the U axis for $T\rightarrow 0$. Inset:
Further zoom on the lowest temperatures, emphasising the growth of the error bar with decreasing temperature.} 
\label{fig:phasediag_lowT}
}
\end{figure}

Before proceeding with 
the interpretation of our results and their implications, we conclude this section
with a detailed analysis of our data in the regime of the lowest temperatures accessible to 
our algorithm.
This is particularly important, because a correct determination 
of the vertex divergences for $T\rightarrow 0$ is crucial for 
answering the questions posed in Sec.\,\ref{sec:intro}. 

We start, thus, assessing the numerical accuracy of our results for the 
first red divergence line in the low-$T$ range ($0.003\overline{3} < T < 0.05 < T_K \sim 0.07 $).
Our results are shown in Fig.\,\ref{fig:phasediag_lowT}, together with the 
corresponding error bars. The latter
were obtained from a Jackknife error analysis\cite{Jackknife}, 
which is described in detail in the 
Appendix C.
From the error bars in the main plot and the inset of Fig.\,\ref{fig:phasediag_lowT}
it can be inferred, that the combined scaling ($\beta^3$ of the CT-QMC sampling and 
$\beta^2$ of the Matsubara frequency box of the vertex function for $\Omega_n=0$)
prohibits us to access temperatures lower than $T=0.003\overline{3}$, therefore 
not yielding any further informative results about the vertex divergences.
However, the numerical precision for $T>0.003\overline{3}$ 
was sufficient to accurately define the low-$T$ behavior. 
In fact, we can compare our data with the dotted gray line, showing a linear 
extrapolation of the divergence line to $T\rightarrow 0$ using 
the (higher) temperatures $T=0.05$ and $T=0.025$.
Even considering the growing error bars, the first divergence line shows
a progressive leftwards deviation from the linear extrapolation when reducing the temperature.
This is evidently completely inconsistent with an infinite value of $\widetilde{U}$ of the 
divergence line endpoint for $T\rightarrow 0$. 

\begin{figure*}[t]
{{\resizebox{8.0cm}{!}{\includegraphics {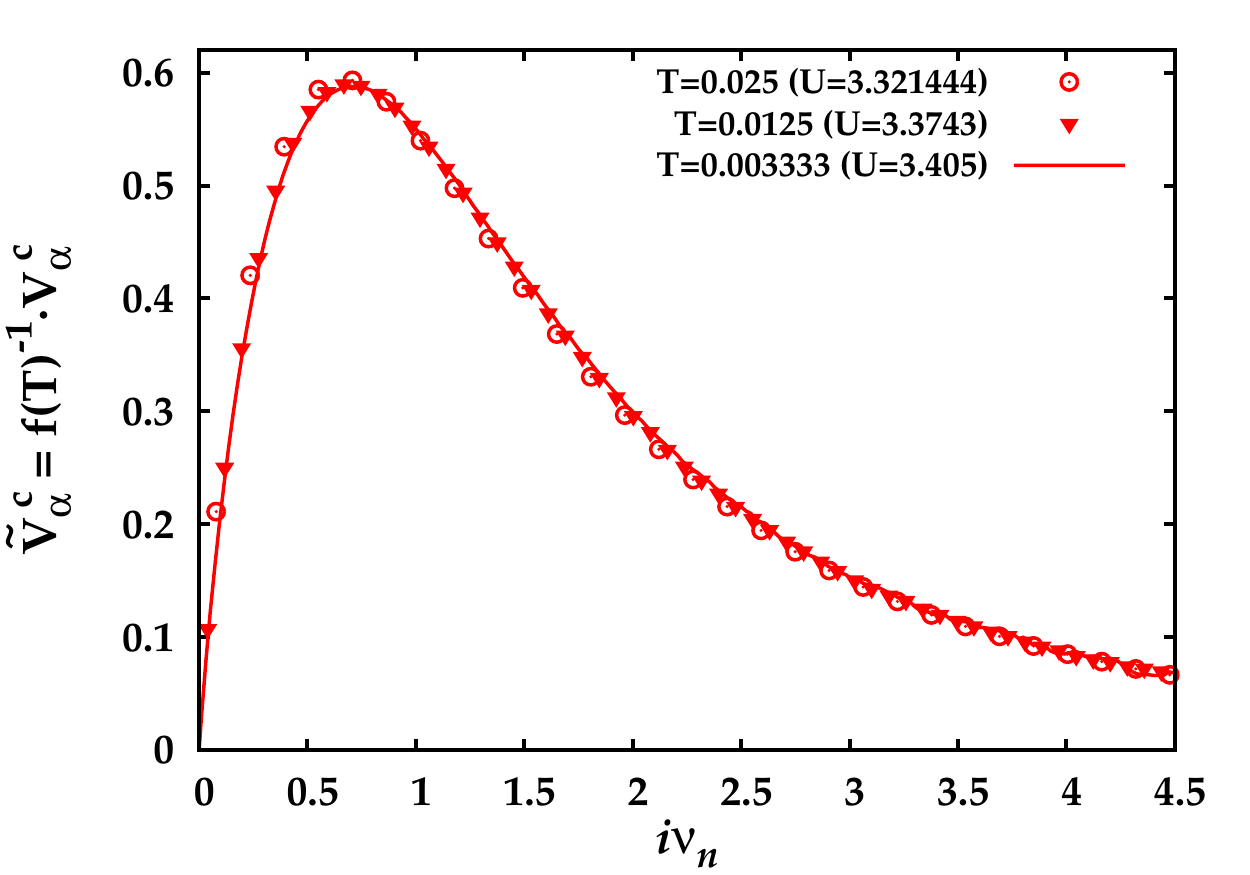}}}}
{{\resizebox{8.0cm}{!}{\includegraphics {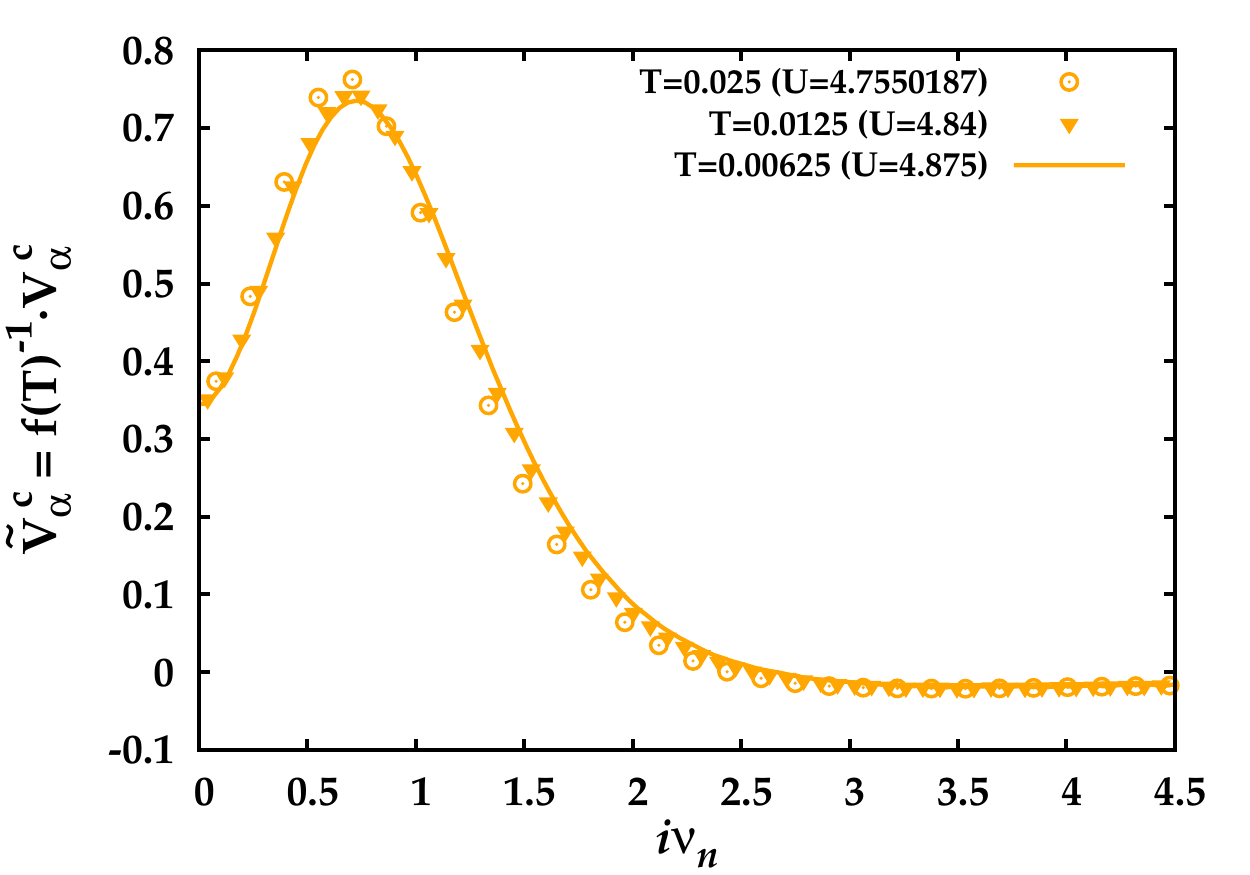}}} 
\caption{Left panel: Eigenvectors of the first red divergence line for a given temperature
multiplied with the inverse of the scaling factor $f(T)$ yielding the eigenvector at $T=0$
governing the frequency structure of the eigenvectors at all temperatures in the $T\ll T_K$ regime. 
Right panel: As left panel, but for the first 
orange divergence line. Here the minor discrepancies can be ascribed to a stronger $U$ dependence of the 
singular eigenvectors in comparison to those of the red lines.}
\label{fig:rescaled_ev}
}
\end{figure*}

That the temperatures considered are low enough to allow for a $T\rightarrow 0$ extrapolation 
is also supported by the behavior of the singular eigenvectors. We discuss here 
the case for the first red and orange divergence, which is representative for all calculated 
divergence lines. 
In fact, for 
$T \ll T_K$ (e.g. for $T \le 0.025$ for the first divergence) the eigenvectors 
do not only display a maximum at a $T$-independent value $i\nu_{n_{\text{max}}}$,
but as functions of $i\nu_n$, they even show a perfect scaling in the 
whole low-$T$ regime (see Fig.\,\ref{fig:rescaled_ev}).
This demonstrates that the low-$T$ frequency structure of the singular eigenvectors, 
and hence, of the vertex divergences, is completely controlled by an underlying, $T$-independent, 
function: $\widetilde{V}^c_{\alpha} (i\nu)$, 
such that $V^c_{\alpha}(i\nu_n, T) = f(T)\widetilde{V}^c_{\alpha} (i\nu)$. 
Our numerical data indicates further that $f(T)$ simply 
represents the conversion factor needed, when taking the $T\rightarrow 0$-limit of the discrete sum
of Matsubara frequencies defining the norm of the eigenvector 
($\sum_{v_n} |V^c_{\alpha} (i\nu_n, T)|^2 = 1$): $f(T) = \sqrt{2\pi T}$. 
In Fig.\,\ref{fig:rescaled_ev} the correspondingly rescaled eigenvectors ($=\widetilde{V}^c_{\alpha} (i\nu)$)
for 
the first red and orange divergence line are shown. These are extracted from the 
data for $V^c_{\alpha}(i\nu_n, T)$ by exploiting the low-$T$ scaling relation:

\begin{eqnarray}
\widetilde{V}^c_{\alpha} (i\nu)=\frac{{V}^c_{\alpha}(i\nu_n,T)}{f(T)} = 
\frac{{V}^c_{\alpha}(i\nu_n,T)}{\sqrt{2\pi T}} 
\end{eqnarray}

In the case of $\widetilde{V}^c_{\alpha}(i\nu)$, $i\nu$ represents continuous imaginary frequencies.  

\section{Discussion and analysis}


Our numerical study of the vertex divergences in the AIM presented
in the previous sections, and the
comparison of the results to the ones of the Hubbard model in DMFT,
yield clear-cut answers to several open questions on this subject, which 
were mentioned at the end of Sec.\,I.

In particular,  the results definitely demonstrate that {\bf(i)} the MIT
does \emph{not} represent  the essential ingredient to induce 
vertex divergences (as well as
the associated non-perturbative  manifestations). 
This is proven by the similarity of the low-$T$  behavior of the vertex
divergence lines in the Hubbard model and the AIM,  ending in both cases
at finite $U$ values in the limit $T \rightarrow 0$, although \emph{no}
MIT occurs in the ground state of the latter.
We must conclude, hence, that the occurrence of a MIT can represent 
a \emph{sufficient}, but not a \emph{necessary},
condition to observe vertex divergences. In this respect we recall that in the
phase diagrams of the Hubbard/FK
models the MIT cannot be reached (from the
non-interacting or the high-$T$ perturbative region) without
crossing (at least) one  divergence line: Only in this somewhat more limited
context, the vertex divergences can be regarded as
''precursors'' of the MIT, as originally proposed\cite{Schaefer2013}.
 
The picture emerging from our AIM data does not contradict, however, the
physical considerations made in Ref.\,[\onlinecite{Gunnarsson2016, Gunnarsson2017}], where one could 
relate the suppression of the charge susceptibility\cite{Gunnarsson2017},
driven by the formation of a local magnetic moment\cite{GunnarNote}, to the onset of 
the divergences. The same physical mechanism can also 
induce, depending on the model, the appearance of a MIT. This scenario would be, thus, coherent with 
our numerical finding of divergence lines without a MIT. 

At the same time, the origin of the rather striking similarity between
the low-$T$ curvature of the divergence lines and the MIT in the DMFT
phase diagram of the Hubbard model can be further rationalized, as we will
discuss below, in terms of the relation to $T_K$.

\begin{figure}[t]
{{\resizebox{8.0cm}{!}{\includegraphics {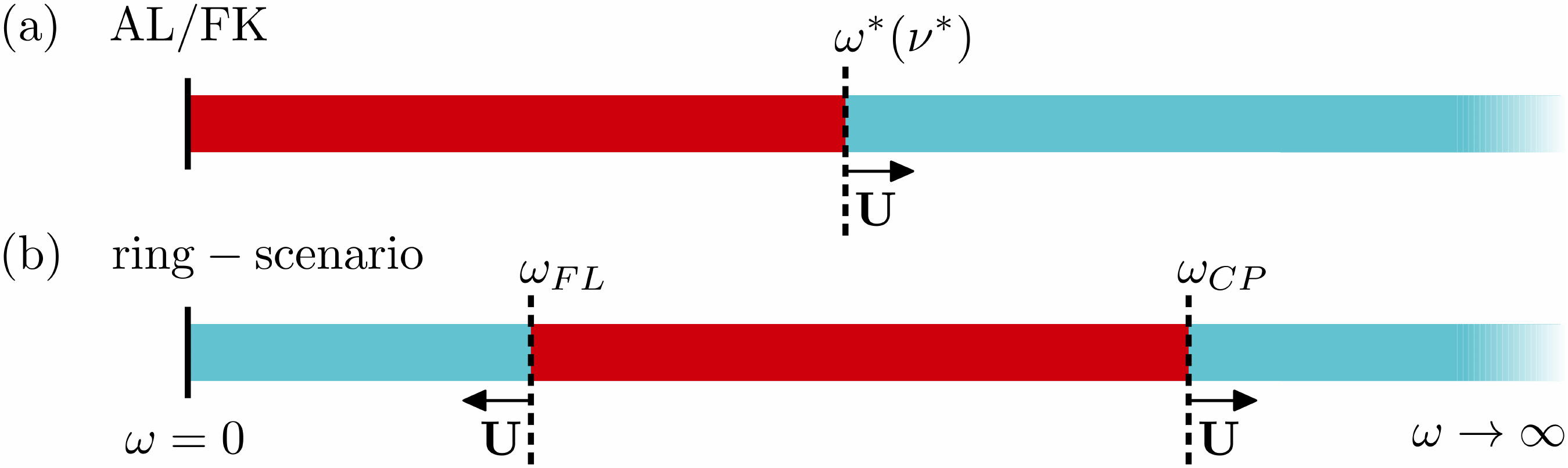}}}
\caption{Schematic representation of the energy scale(s) controlling the border(s) 
of the perturbative/non-perturbative regimes, (a): in the AL/FK case, which is relevant
for the large $T$ and $U$ limit of the AIM and the DMFT solution of the Hubbard model, (b): 
for the "ring-scenario" which was proposed in Ref.\,[\onlinecite{Schaefer2016}], for 
the Hubbard model in the correlated metallic regime (see text). Note that for the AIM 
with a large bandwidth, as studied here, only one energy scale exists.} 
\label{fig:ring_scenario}
}
\end{figure}

Further, {\bf(ii)} the observation of an overall qualitatively
similar $T$--$U$ diagram (down to very low $T$) rules out the proposed
explanation of the line re-entrance shape in terms of the so 
called "ring-scenario"\cite{Schaefer2016}. This 
proposed scenario represents a simple
generalization of the theoretical framework applicable at high $T$
and large $U$.
We recall\cite{Schaefer2016} that, in the latter regime, the existence of infinitely many
divergence lines could be interpreted as a direct manifestation of an underlying 
energy scale $\nu^*$ in the
Matsubara frequency space: A divergence occurs whenever a Matsubara frequency  
($\nu_n =\frac{\pi}{\beta} (2n +1)$) equals $\nu^*$. While it was already made
clear in previous studies\cite{Schaefer2016}, that this high-$T$
explanation does not work at low-$T$, a simple
generalization was proposed: At low-$T$ \emph{two} underlying energy scales might
control the vertex divergences. This idea would have matched well previous
results showing that in the DMFT solution of the Hubbard model (for
roughly the same interaction values where the divergences
occur) \emph{two} energy scales $\omega_{FL}$ and $\omega_{CP}$ appear in
the low-energy sector\cite{Byzuk2007}, where the renormalized quasiparticle
excitations of the systems are defined. From a merely theoretical
viewpoint, the
separation of low-energy scales can be ascribed\cite{Held2013} to the 
self-consistent renormalization of the 
electronic
bath of the auxiliary AIM of DMFT in the correlated metallic regime. This has
also important observable consequences, like the
emergence of kinks in the spectral functions\cite{Byzuk2007} and the specific 
heat\cite{Toschi2009} of
the Hubbard model. In the perspective of the vertex divergences, the
emergence of two energy scales on the real axis would correspond to a
situation (referred to  as ``ring-scenario'') where the perturbative physics is preserved not 
only -- as usual -- at high energy (for $\omega > \omega_{CP}$), but also in the
lowest frequency Fermi-Liquid $(FL)$ regime (for $\omega < \omega_{FL}$), for a schematic representation 
see Fig.\,\ref{fig:ring_scenario}.
The non-perturbative effects, instead, would appear first at 
intermediate energies, i.e., for  $\omega_{FL}  < \omega <  \omega_{CP}
$, and reach the Fermi level, only at the MIT. 

The existence of two scales could be indeed reflected in the
observed more complex non-local structure of the singular eigenvectors in
Matsubara space. Moreover, this interpretation would also have the appealing advantages of
providing a one-to-one correspondence of the vertex divergences to
physical observables (the kinks), and, at the same time, of avoiding the
necessity to deal with a Fermi-Liquid ground state, of intrinsic
non-perturbative nature.

Our finding of qualitatively similar divergence lines, also in the low-$T$ area of the
$T$--$U$ diagram of the AIM, however, makes a general validity of the proposed
``ring-scenario'' very unlikely: In our AIM with a fixed and large conduction electron bandwidth, 
the intrinsic origin of
the separation of energy scales (i.e., the self-consistent
renormalization of the electronic bath) is missing, see the discussion in 
Ref.\,[\onlinecite{Held2013},\onlinecite{Kainz2012}].

As a consequence, from a theoretical point of view, one will indeed face  
the challenge of reconciling the observation of well defined
Fermi-Liquid properties at low-energies with the evident breakdown
of perturbation theory marked by the multiple divergence lines. 
In other words, it will be necessary to describe and fully understand the emergence of an intrinsically 
non-perturbative Fermi-Liquid phase. 

From a more physical point of view, the similarity of the divergence lines in the Hubbard model 
and the AIM also excludes a direct connection of the divergences to the kinks in the self-energy, 
which are present\cite{Byzuk2007, Toschi2010} and absent\cite{Kainz2012, Held2013} in the 
two respective models. Note however, that the kinks in the DMFT solution of the Hubbard model
are also related to the Kondo temperature\cite{Held2013}, 
so there might be an indirect connection as for the 
bending of the divergence lines around $T_K$, see {\bf (iv)}.

{\bf(iii)} The presence of several distinct intersections of the
divergence lines with the $T=0$ axis poses the question as to whether irreducible vertex divergences
occur also on the
real frequency axis. In fact, this behavior is radically different
from the one found in the FK model\cite{Schaefer2016,Janis2014,Ribic2016}. In the
latter case, all divergences lines accumulate at $T=0$ for a
non-zero interaction value\cite{Schaefer2016,Ribic2016} ($U=D/\sqrt{2}$), which corresponds to the
unique vertex divergence on the real frequency
axis\cite{Schaefer2016,Janis2014}.
The low-$T$ spread displayed by the divergence lines of the AIM is,
thus,  fully incompatible with the simpler FK scenario of a single divergence
point of the vertex in real-frequency. In this respect, important insight is provided by the analysis
of the temperature evolution of the singular eigenvectors $V^c_{\alpha}(i\nu_n)$ for the
different lines, especially in the low-$T$ regime where their behavior
can be fully described by a rigid scaling (see Sec.\,\ref{sec:lowT_red1}  and
Fig.\,\ref{fig:rescaled_ev}). In fact, close to the divergence, $\Gamma_c$ can be
approximated as in Eq.\,(\ref{eq:divapprox}).
The scaling properties of the eigenvectors,
combined with the different (odd/even) symmetries (under $\nu_n
\rightarrow -\nu_n$) of the eigenvectors corresponding to the red/orange divergences, 
provide a strong indication that real-frequency divergences in the limit of zero
frequencies, i.e. $i\Omega_n=0,i\nu_n,i\nu_{n'}\rightarrow 0$, will appear at the
endpoints of \emph{all} orange lines. 
This is due to the fact that for the rescaled eigenvectors of orange divergence lines the
value of the lowest frequency component of the singular eigenvector
 is finite and non-zero in the $T \rightarrow
0$ limit 
($\widetilde{V}^c_{\alpha}(i\nu=0) \neq 0 $, see Sec.\,\ref{sec:lowT_red1} 
and Fig.\,\ref{fig:rescaled_ev}). 
On the contrary,   
the vanishing of $\widetilde{V}^{c}_{\alpha}(i\nu=0)$ for red divergences, 
which is enforced by the odd symmetry of the eigenvector, suggests the
absence of similar real-frequency divergences as for the orange
endpoints. 
This would represent a 
further, major differentiation between
the two kinds of divergences, as only the orange ones would be mirrored
by corresponding divergences at the origin of the real-frequency axis at the $T=0$ endpoints.
Of course, our analysis can not exclude, that the real frequency vertex functions might 
display additional divergences at finite non-zero frequencies.

{\bf(iv)} The precise determination of the Kondo temperature in the
$T$--$U$ diagram of the AIM we considered (Fig.\,\ref{fig:kondo_phasediag}) provides novel insights
into the problem of the vertex divergences in the
correlated metallic regime. 
First, as we briefly mentioned in Sec.\,III, the relatively
featureless and smooth behavior of all divergence lines for $T < T_K$
indicates that unexpected bending below the lowest
temperature where the QMC calculations of the impurity vertex
functions are feasible, is highly improbable and also incompatible with the perfect scaling 
of the singular eigenvectors discussed in Sec.\,\ref{sec:lowT_red1}. This represents an important,
physics-based argument supporting all low-$T$ results and
considerations discussed before. 
Second, we must recall that the Kondo screening in the AIM  is not taking place
as a sharp transition. On the contrary, it is known that the impurity 
magnetic moment screening starts to 
become
progressively effective at temperature larger than $T_K$. 
In fact, considering the low-energy spectral properties of the impurity site,
a natural estimate of the crossover temperature, leads to values more 
than five times larger than the ``standard'' $T_K$ (see Sec.\,\ref{sec:tudiagram}).
This is relevant for the interpretation of our results, as the bending
of the divergence lines, and in particular their re-entrance behavior,
is taking place in this crossover regime. Consistently, in the same parameter region, the
qualitative change in the structure of the eigenvectors for the red lines (and
of the corresponding divergence of $\Gamma_c$ from
localized to non-localized in frequency space) takes place.  
The emerging scenario for the breakdown of perturbation expansion in
the AIM is, thus, the following: Vertex divergences  with a
relatively simple structure (straight linear behavior of all
divergence lines, fully localized eigenvectors, etc.) are associated to the formation of a local moment,
and to the related net separation of energy scales, marked by a spectral
gap between the Hubbard bands. This agrees with our understanding in the 
AL\cite{Schaefer2013, Rohringer_Diss, Schaefer2016, Rohringer_AL}.

The progressive Kondo screening
occurring by lowering $T$ towards $T_K$ is responsible for the
gradual, but important deviations from this 
``simpler'' divergent behavior. 
In particular, it is interesting to note that the screening of
the local moment appears, to some extent, to ''counter'' the breakdown of
the perturbation expansion (low-$T$ re-entrance) of the lines. This
effect, however, is only partial, as the divergence lines for $T < T_K$ 
are not bending up to $U =
\infty$, but rather display multiple (presumably infinite) endpoints on
the $T=0$ axis.

{\bf(v)} From the numerical results obtained for the AIM and from the
considerations made hitherto, it is possible to
formulate specific, though somewhat heuristic, predictions for the structure of the
divergence lines in the DMFT solution of the Hubbard model. In particular, we will focus on the
most interesting regime of the coexistence region surrounding the
Mott-Hubbard MIT. In fact, due to the considerable accumulations of
vertex divergences close to the MIT, no detailed study of the
irreducible vertex functions in this parameter region has been performed so far. 

The starting point for our prediction is the connection discussed
above {\bf (iv)}, between the Kondo screening, controlled by $T_K$,
and the bending of the divergence lines.  Further, it is also important that for $ T < T_K$ each
of the (infinitely) many lines smoothly continues down to their respective finite endpoint on the $T=0$
axis {\bf (iii)}. In fact,  a converged DMFT
solution is defined through the self-consistently determined
electronic bath of the auxiliary AIM. This means that the AIM of  a given DMFT
solution would be characterized by a $T_K$ depending not only on $U$ 
but also on $T$, $n$, etc.  
Such an ``effective'' $T_K^{\text{DMFT}}(U,T)$ becomes zero at the Mott-Hubbard
MIT at $T=0, U \rightarrow U_{c2}$. This suggests that, the divergence 
lines appearing in the  $T< T_K$
part of the whole $T$-$U$ parameter space of the AIM will be 
squeezed in the region $U< U_{c2}$ of the DMFT solution
of the Hubbard model. This will evidently result in an
accumulation of divergence lines close to $U_{c2}$. 
For  $T \neq 0$, instead,  one will cross, by increasing $U$, a first
order MIT where a discontinuity of the physical properties occurs. 
Such a discontinuity affects, among other observables like the double
occupancy or the kinetic energy, \emph{also} the self-consistently determined
electronic bath of the auxiliary AIM. Hence, one must expect that the
number of negative eigenvalues of $\chi_c$  (and thus, of divergences lines
already crossed) will be \emph{different} on the two sides of the
transition, except for $T=0$ and at the critical endpoint where the
MIT is continuous.  More specifically, we recall that the Mott insulating phase is characterized by an 
essentially
unscreened local moment. As a consequence, on the insulating side of
the MIT we will observe only very minor corrections w.r.t. the divergence
lines computed in the AL (straight lines with frequency
localized red divergences, etc.). On the metallic side of the MIT,
instead, the electronic bath is associated with a finite
$T_K^{\text{DMFT}}(T,U)$, and, thus, the screening effect, partially mitigating the
vertex divergences, will be at work. As a result, for a fixed $T \neq
0$ one would find here a fewer number of negative eigenvalues
(and, hence, of divergences lines) than on the insulating side. 
Eventually, for low-enough $T \sim T_K^{\text{DMFT}}(U,T)$, the divergence lines on the
metallic side will show the typical bending behavior, and display
very frequency delocalized eigenvectors.

\section{Conclusion and Outlook}

In this work we have studied the divergences of the irreducible vertex
functions occurring in an Anderson impurity model with a fixed electronic
bath, aiming at gaining novel insights about the breakdown of
many-body perturbation theory in correlated metallic systems.
In fact, the numerical solution of the AIM, computed at
high accuracy in CT-QMC, fully captures the physics of low-temperature
quasi-particle excitations, as those observed in correlated metals. 
It avoids, however, the additional complication of a
self-consistent adjustment of the electronic bath required by DMFT
calculations. Hence, the AIM represents a fundamental test bed to address
several open issues posed in the literature, about the
different non-perturbative manifestations in quantum many-body theory.

Indeed, our study could clarify a set of relevant
questions about the
interpretation and the consequences of the divergences of the two-particle irreducible
vertex functions. In particular, our results rule out 
that the Mott-Hubbard transition plays a crucial role as the origin of the multiple divergence lines.
This limits the previously proposed interpretation\cite{Schaefer2013} of the vertex
divergences as ``precursors'' of the MIT in the sense of a necessary condition for 
vertex divergences to occur, consistently with the physical interpretation presented in 
Refs.\,[\onlinecite{Gunnarsson2016, Gunnarsson2017}]. 
By a thorough analysis of the
low-temperature sector, we could ascribe, at the same time, important
characteristics of the vertex divergences, such as their structure in
Matsubara frequency space and the re-entrance of the divergences line, to the
screening processes of the local magnetic moment occurring when
approaching  $T_K$. Moreover, our data for $T\ll T_K$ has unveiled a perfect
scaling of the singular eigenvectors, allowing us to 
extrapolate the $T=0$ behavior of the vertex divergences on the
real-frequency axis. Finally, exploiting the insights gained from the low-$T$ analysis, we could propose 
a heuristic prediction about the vertex divergences
in the particularly relevant case of the coexistence region in
the DMFT phase diagram of the Hubbard model. 

Having clarified important properties of the vertex
divergences in the correlated metallic phase (partly with an unexpected outcome which corrected previously
made assumptions) our work also demonstrates how valuable studies are,
where the many-body correlation effects are realized in the most
fundamental fashion. This suggests future extensions of this
work, by considering systems with embedded clusters of impurities, to
investigate the role played by short-range correlations in the
breakdown of the  quantum many-body perturbation expansion.

\section*{Acknowledgements}

We thank O. Gunnarsson, G. Sangiovanni,
G. Rohringer, M. Capone, P. Thunstr\"om, D. Springer, A. Hausoel, L. Del Re, 
and S. Ciuchi for insightful discussions.
We acknowledge support from the Austrian Science Fund
(FWF) through the DACH Project No. I 2794-N35 (P.C., T.S.), and the
 SFB ViCoM Project No. F41 (A.T., K.H.) and from the Vienna Scientific Cluster Research Center 
 funded by the 
 Austrian Federal Ministry of Science, Research and Economy (bmwfw) (P.G.). 
 T.S. has also received funding from the European Research Council under the European 
 Union Seventh Framework Program (Program No. FP7/2007-2013) ERC Grant Agreement No. 306447.
 Calculations were performed on the Vienna Scientific Cluster (VSC).

\section*{Appendix}

\subsection*{Appendix A: The atomic limit and the inflection point of $\bm{\mathrm{Im}\, G(i\nu_n)$}}

\begin{figure}[tb]
{{\resizebox{8.5cm}{!}{\includegraphics {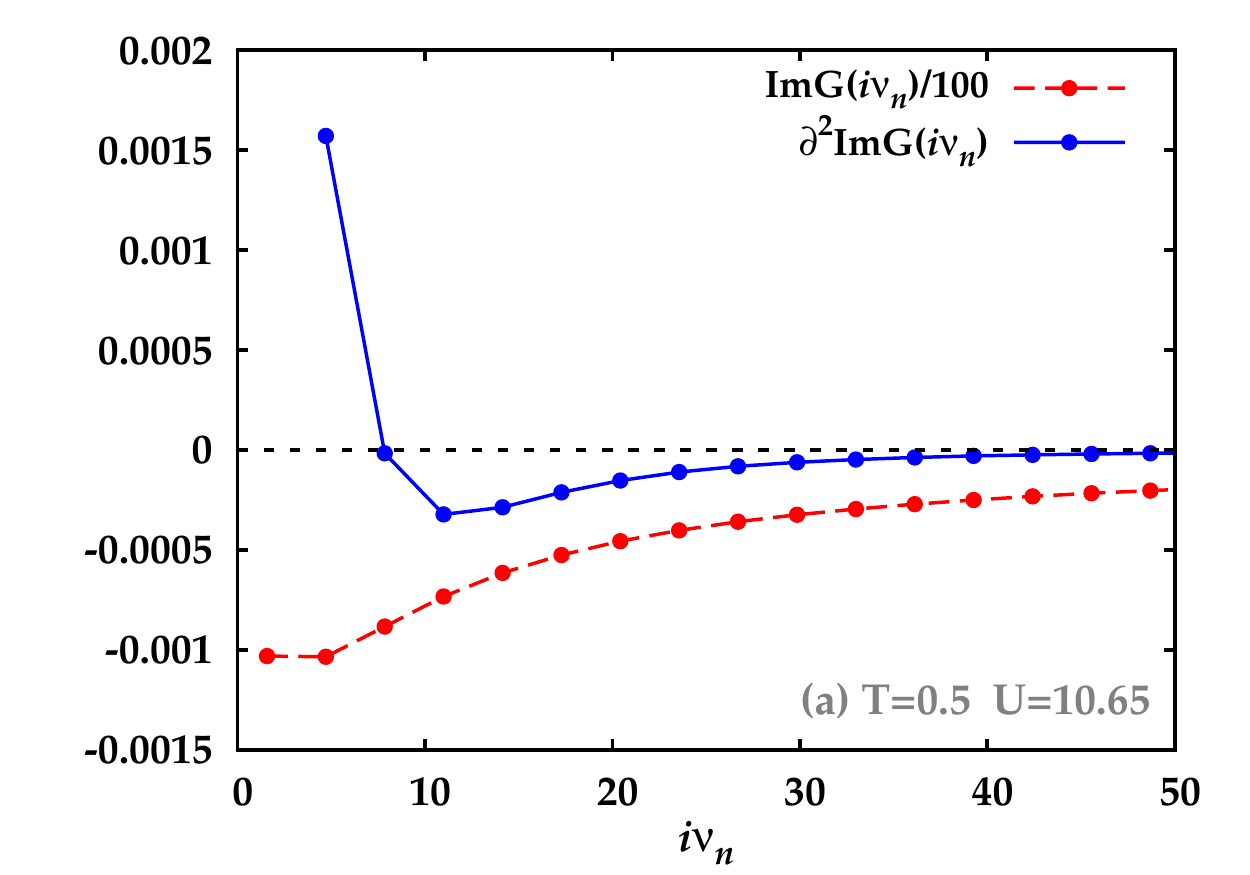}}}}
{{\resizebox{8.5cm}{!}{\includegraphics {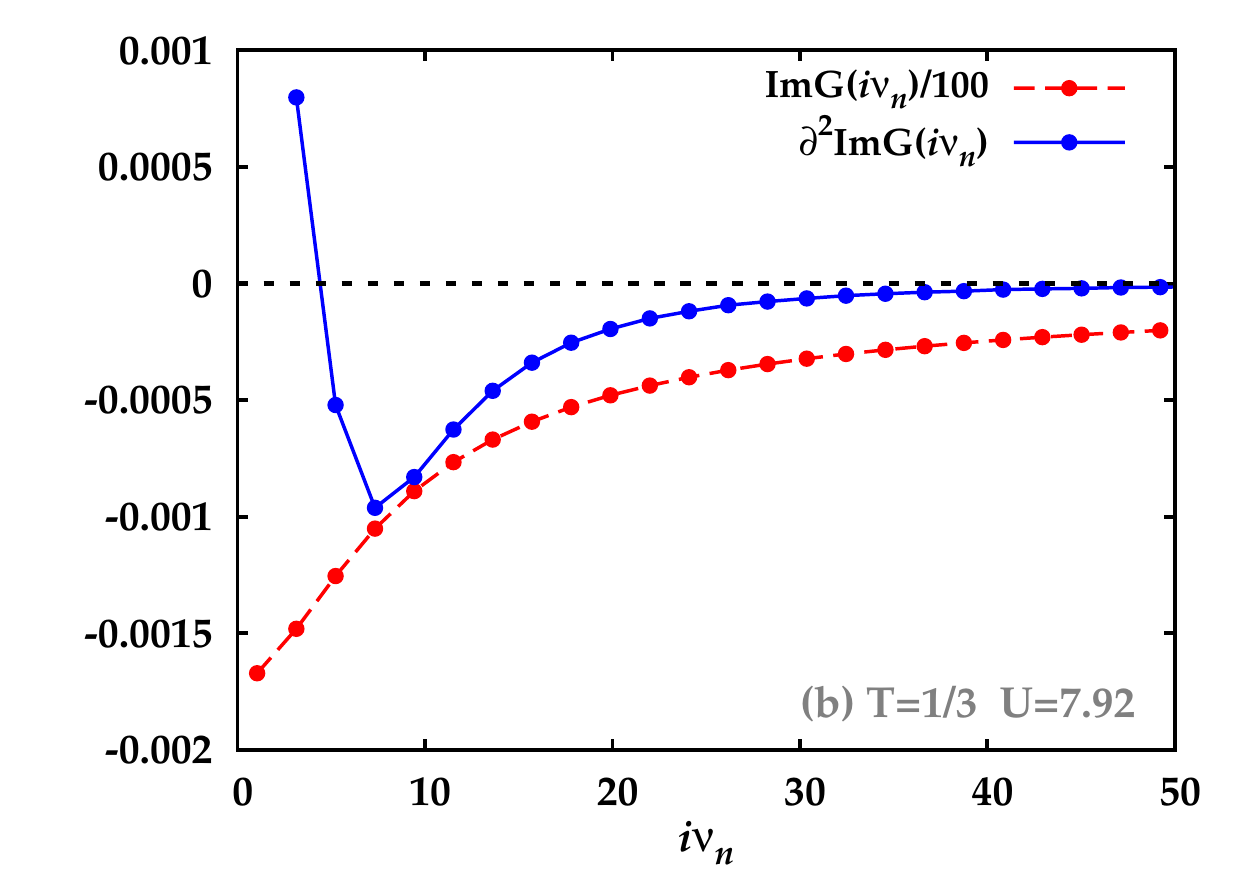}}}}
{{\resizebox{8.5cm}{!}{\includegraphics {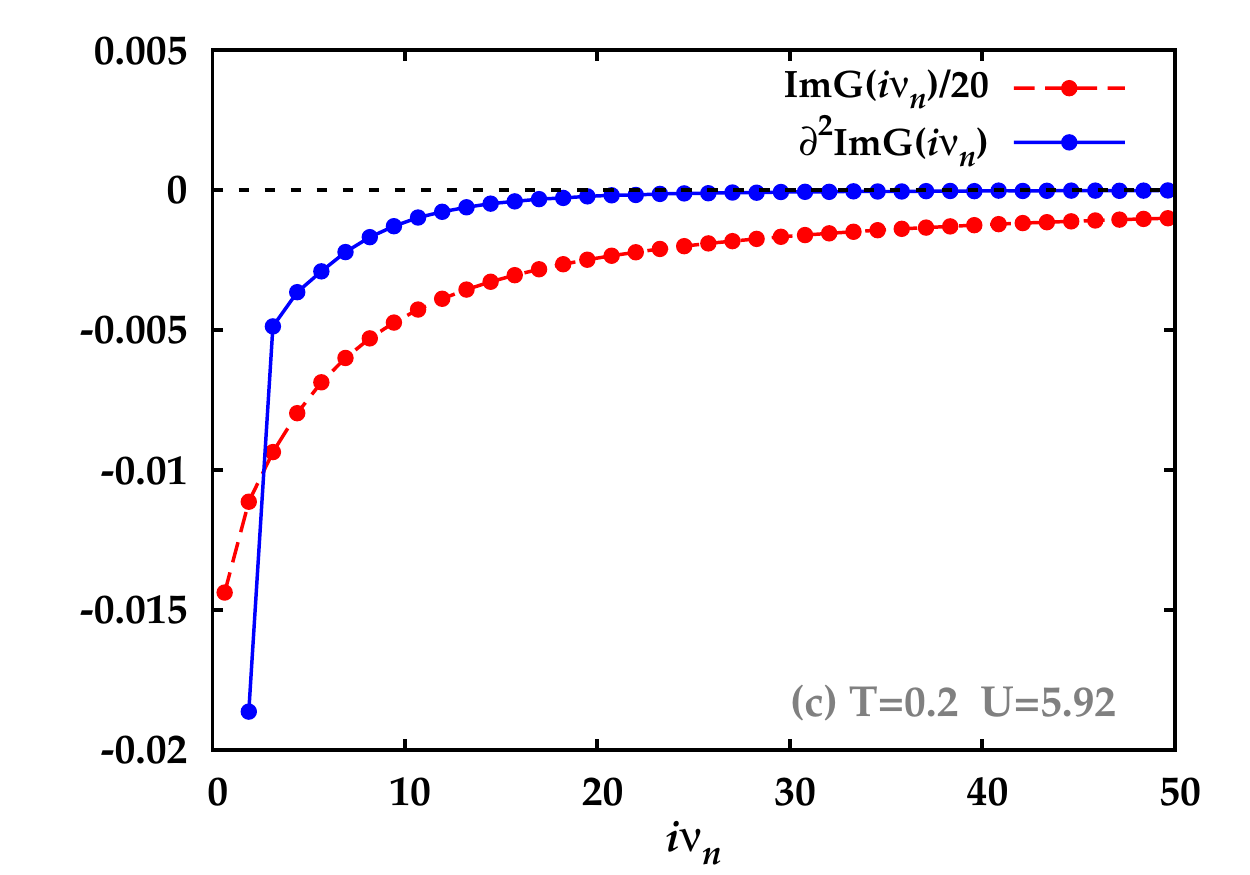}}}
\caption{Comparison of $\mathrm{Im}\, G(i\nu_n)$ (rescaled by a proper factor) and 
$\partial^2\mathrm{Im}\, G(i\nu_n) = \frac{\partial^2\mathrm{Im}\, G(i\nu_n)}{\partial \nu_n^2}$ 
for three different temperatures along the third red divergence line of the AIM.} 
\label{fig:appendix_a_l5_b2}
}
\end{figure}

As mentioned in Sec.\,\ref{sec:intro}, in the case of the atomic limit, 
the inflection point of 
$\mathrm{Im}\, G(i\nu_n)$ is found\cite{Rohringer_AL} at the frequency 
$\nu_n=\nu^{\ast}$, i.e. the energy scale which is also 
governing the position of the divergence lines and the frequency structure of the 
localized singular eigenvectors corresponding to red divergence lines. 
As we briefly discuss in Sec.\,\ref{sec:tudiagram},
this connection is also found for the Anderson impurity model, in the regime of high-$T$ and
large interaction. The results are shown explicitly in 
Fig.\,\ref{fig:appendix_a_l5_b2} of this Appendix, 
where the rescaled $\mathrm{Im}\, G(i\nu_n)$ 
is compared to $\partial^2\mathrm{Im}\, G(i\nu_n) = 
\frac{\partial^2\mathrm{Im}\, G(i\nu_n)}{\partial \nu_n^2}$ 
for the third red divergence line. Note that in the simple method used to compute the 
second derivative of $\mathrm{Im}\, G(i\nu_n)$, namely finite differences, 
there is no data for the first frequency.
For $T=0.5$ (see Fig.\,\ref{fig:appendix_a_l5_b2}a), it is evident 
that the inflection point is located at the third Matsubara frequency, which 
agrees with the atomic limit description.  With decreasing temperatures, the 
inflection point moves from the third frequency towards lower frequencies, which is visible in 
Fig.\,\ref{fig:appendix_a_l5_b2}b, where it is found between the second and third frequency for 
$T=0.3\bar{3}$. Although the divergence line shows still a rather linear behavior (see main part, 
Fig.\,\ref{fig:phase_diag_all}), the inflection 
point completely disappears for $T=0.2$ (see Fig.\,\ref{fig:appendix_a_l5_b2}c).

\subsection*{Appendix B: The numerical extraction of the Kondo temperature $\bm{T_K}$}

\begin{figure}[tb]
{{\resizebox{8.5cm}{!}{\includegraphics {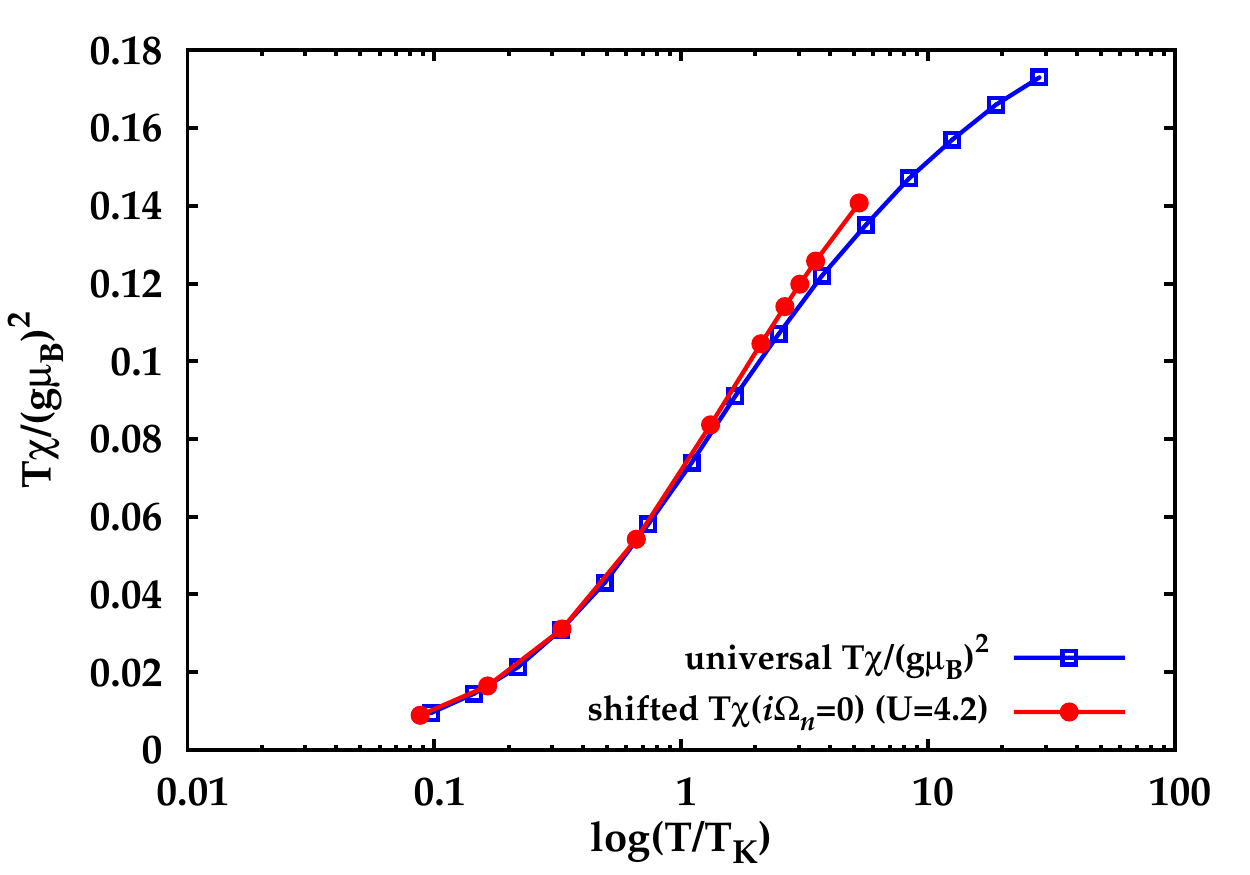}}}
\caption{The shifted data for $T\chi_s(\omega=0)$ for several temperatures at $U=4.2$, compared
to the universal result for $T\chi_s(T)$ (see Ref.\,[\onlinecite{Krishna}]).} 
\label{fig:appendix_b_Chi}
}
\end{figure}

To extract the Kondo temperature $T_K$ numerically the static local magnetic susceptibility of the impurity
$\chi_s(i\Omega_n=0)$ has been calculated by integrating 
$\chi_s(\tau)=g^2\langle S_z(\tau) S_z(0)\rangle$, with $g=2$ 
(computed with w2dynamics) over the interval $[0,\beta]$.

\begin{equation}
\chi_s(i\Omega_n=0) = \int \limits_0^\beta \chi_s(\tau)
\end{equation}
which corresponds to its Fourier transform for $i\Omega_n=0$. 
The data for $\chi_s(i\Omega_n=0)$ has been computed for several $T$, and then 
compared to the universal result for $\frac{T\chi_s(T)}{(g\mu_B)^2}$, 
computed for the spin-$\frac{1}{2}$ Kondo Hamiltonian 
in Ref.\,[\onlinecite{Krishna}] (cf. also Ref.\,[\onlinecite{Wilson_kondo}]), 
where $\mu_B$ is 
the Bohr magneton. 
Plotted as a function of $log(T/T_K)$ the Kondo temperature can thus be obtained with high precision
by shifting the numerical data for $T\chi_s(i\Omega_n=0)$ 
onto the universal result (on a logarithmic $T$-scale). For the case of $U=4.2$ the shifted results
are shown in Fig.\,\ref{fig:appendix_b_Chi}. The data for $T_K$ obtained through this procedure 
are the blue crosses reported in Fig.\,\ref{fig:kondo_phasediag} of the main paper.

\subsection*{Appendix C: The determination of $\bm{\widetilde{U}}$ and the associated error bars}

\begin{figure}[tb]
{{\resizebox{8.5cm}{!}{\includegraphics {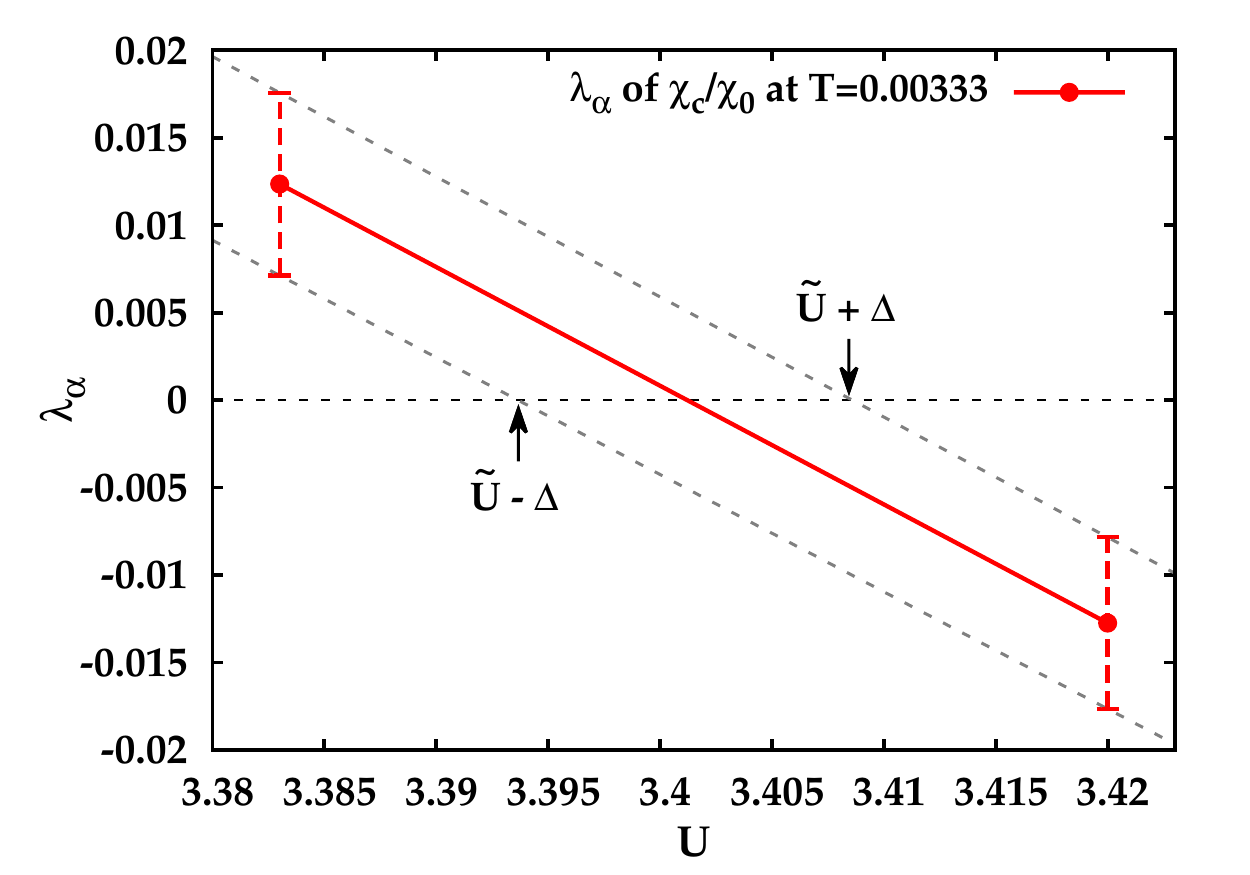}}}
\caption{The singular eigenvalue $\lambda_{\alpha}$ of 
$(\chi_c/\chi_{ph,0})^{\nu_n \nu_{n'} \Omega=0}$ for
$T=0.003\bar{3}$ and two different $U$ values ($3.383$, $3.42$) is shown. A linear interpolation is used for the 
determination of $\widetilde{U}$ (red solid line). The Jackknife error analysis provided error bars 
for the two results for $\lambda_{\alpha}$ (red dashed), which are, then, used to estimate the 
error bar of $\widetilde{U}$ (gray dashed lines). } 
\label{fig:appendix_c_Jackknife}
}
\end{figure} 

To estimate $\widetilde{U}$ for a given temperature 
at least two separate calculations of $\chi_c^{\nu_n \nu_{n'} \Omega_n=0}$ 
are necessary. As an example of the procedure, we show the case of the first divergence line
for $T=1/300 \approx 0.00333$. 
For this case the smallest eigenvalue of $(\chi_c/\chi_{ph,0})^{\nu_n \nu_{n'}}$ 
(see Sec.\,\ref{sec:calculation_ctqmc}) is plotted as a function of $U$, 
see Fig.\,\ref{fig:appendix_c_Jackknife}. 
This, in turn, allows us to adopt an interpolation or extrapolation procedure for an 
estimation of $\widetilde{U}$, 
depending on whether we find two eigenvalues 
with different signs or not.
In the following a bisection procedure is used until $\widetilde{U}$ is known to the desired 
accuracy for the given temperature. 
In the calculations shown throughout this work, we have performed calculations 
until we reached an interval in $U$ of the order of $\mathcal{O}(10^{-1})$ to $\mathcal{O}(10^{-3})$. 
For the first line and high-$T$ the coarser interval was used, for the subsequent lines a more 
refined interval was employed, as well as for the low-$T$ results of all lines.  
Each calculation was performed on the Vienna Scientific Cluster (VSC3) 
using about $10.000$ to $15.000$ CPU hours, in the case of the 
low-$T$ calculations $25.000$ ($\beta=160$) or $50.000$ ($\beta=300$) CPU hours were used. 

Finally, to 
estimate the error bars of the interaction values $\widetilde{U}$ it is necessary to extract, first, 
the error of the eigenvalues of $(\chi_c/\chi_{ph,0})^{\nu_n \nu_{n'} \Omega_n=0}$. To this end a $(n-1)$ Jackknife 
method\cite{Jackknife} was used. 
Specifically, with the same CPU hours used for the calculations discussed above, 
$25$ bins were produced. From these $25$ different results 
for $(\chi_c/\chi_{ph,0})^{\nu_n \nu_{n'} \Omega_n=0}$
$26$ eigenvalues were produced, according to the ($n-1$) Jackknife method\cite{Jackknife}, 
which also 
provides an expression for the standard deviation. 
As a last step, the intersection points of the interpolation of the 
maximum and minimum error of the 
eigenvalues and zero, were used as an estimate for the error bar of $\widetilde{U}$ for a given $T$. 
This procedure 
is also shown explicitly in Fig.\,\ref{fig:appendix_c_Jackknife}.


\begin{references}
\bibitem{ED_Check} We have used the implementation described in the Appendix 
of Ref.\,[\onlinecite{DGA}] with $n_s=5$ and
$T=0.08$.
\end{references}
\end{document}